\documentstyle[preprint,tighten,aps]{revtex}    % Specifies the document style.
\begin{document}           %
\draft
\preprint{\vbox{\noindent
Submitted to Physical Review D\hfill INFNFE-3-94\\
          \null\hfill  INFNCA-TH-94-10}}
\title{Neutrinos from the Sun: \\
       experimental results confronted with solar models
      }
\author{
         V.~Castellani$^{(1)}$,
         S.~Degl'Innocenti$^{(2,3)}$,
         G.~Fiorentini$^{(2,3)}$,
         M.~Lissia$^{(4)}$
        and B.~Ricci$^{(3,5)}$
       }
\address{
$^{(1)}$Dipartimento di Fisica dell'Universit\`a di Pisa, I-56100 Pisa, \\
Osservatorio Astronomico di Collurania, I-64100 Teramo, and Universit\`a
dell'Aquila, I-67100 L'Aquila \\
$^{(2)}$Dipartimento di Fisica dell'Universit\`a di Ferrara, I-44100 Ferrara \\
$^{(3)}$Istituto Nazionale di Fisica Nucleare, Sezione di Ferrara,
        I-44100 Ferrara \\
$^{(4)}$Dipartimento di Fisica dell'Universit\`a di Cagliari, I-09100
        Cagliari, \\ and Istituto Nazionale di Fisica Nucleare, Sezione
        di Cagliari, I-09100 Cagliari \\
$^{(5)}$Scuola di Dottorato dell'Universit\`a di Padova, I-35100 Padova.
        }
\date{\today}
\maketitle                 % Produces the title.
\begin{abstract}
For standard neutrinos,
recent solar neutrino results together with the assumption of a
nuclearly powered Sun imply severe constraints on the individual
components of the total neutrino flux:
$\Phi_{\text{Be}}\leq 0.7 \times 10^9 \text{cm}^{-2}\text{s}^{-1}$,
$\Phi_{\text{CNO}}\leq 0.6 \times 10^9 \text{cm}^{-2}\text{s}^{-1}$, and
$64 \times 10^9 \text{cm}^{-2}\text{s}^{-1} \leq
\Phi_{pp+pep} \leq 65 \times 10^9 \text{cm}^{-2}\text{s}^{-1}$
(at $1 \sigma$ level). The bound on $\Phi_{\text{Be}}$
is in strong disagreement with the
standard solar model (SSM) prediction
$\Phi_{\text{Be}}^{\text{SSM}}\approx 5\times 10^9\text{cm}^{-2}\text{s}^{-1}$.
We study a
large variety of non-standard solar models with low inner
temperature, finding that the temperature profiles $T(m)$ follow the
homology relationship: $T(m)=kT^{\text{SSM}}(m)$, so that they are specified
just by the central temperature $T_c$. There is no value of $T_c$ which can
account for all the available experimental  results.
Even if we only consider the Gallium and Kamiokande
results, they remain incompatible.
Lowering the cross section $p+{}^7\text{Be}\to \gamma+ {}^8\text{B}$
is not a remedy. The shift of
the nuclear fusion chain towards the $pp$-I termination could be
induced by a hypothetical low energy resonance in the
$^3\text{He}+ {}^3\text{He}$
reaction. This mechanism gives a somehow better, but still bad
fit to the combined experimental data. We
also discuss what can be learnt from new generation experiments,
planned for the detection of monochromatic solar neutrinos, about
the properties of neutrinos and of the Sun.
\end{abstract}
\narrowtext

\section{Introduction}
\label{intro}
The aim of this paper is to examine whether there is still room for an
astrophysics and/or nuclear physics solution of the solar neutrino
problem, in the light of the most recent results of the Gallium
experiments~\cite{Gallex,SAGE}.

We shall demonstrate that these results, when combined with the
information arising from the Chlorine~\cite{Davis} and Kamiokande~\cite{Kamio}
experiments and -- most important -- with the assumption of a
nuclearly powered Sun, severely constrain the individual
components of the solar neutrino flux, under the hypothesis of
standard (zero mass, no mixing, no magnetic moment \ldots) neutrinos.

The arguments leading to these constraints, already outlined
in a previous paper~\cite{Cast94},
are essentially independent of solar models.
The basic assumption concerning the Sun is that the {\em present} total
neutrino flux can be derived from the {\em presently} observed value of
the solar constant. We remark that these constraints have became much more
stringent after the recent
reports from Gallex and Sage~\cite{Gallex,SAGE}.

 For standard neutrinos, these
results provide evidence that the nuclear energy production chain,
see Fig.~\ref{fig1}, is extremely shifted towards the $pp$-I termination and,
as a consequence, the fluxes of $\nu_{\text{Be}}$ and $\nu_{\text{CNO}}$
are strongly reduced with
respect to the predictions of standard solar models.

The situation is the following: i) we can now compare theory and
experiment at the level of individual fluxes, ii) the solar neutrino
problem, i.e. the discrepancy between experimental results and
standard solar models, affects now also the $^7\text{Be}$-nuclei production,
and not only  the rare $^8\text{B}$ neutrinos.

Next, we ask ourselves whether the solar neutrino problem is
restricted to standard solar models. In this spirit, we analyze several
non-standard solar models with an enhanced $pp$-I termination. The
main inputs of any solar model are listed in Table~\ref{SMpar}.
We are aware of
just two ways for enhancing the $pp$-I termination acting on these
inputs:

\noindent i)
adjusting the parameters which affect the inner solar temperature,
so as to build low inner-temperature solar models,

\noindent ii)
adjusting the $^3\text{He}$ nuclear cross sections.

\noindent
We note that the $p+{}^7\text{Be}\to \gamma+ {}^8\text{B}$
cross section does not influence the $pp$-I branch.

As a relevant and common feature of all the low-inner-temperature models, we
find a homology relation for the temperature profiles,
$T(m)=kT^{\text{SSM}}(m)$, where $k$ depends on
the input parameters,
but it is independent of the mass coordinate $m$ in the inner radiative
zone (at least for $m=M/M_0<0.97$), and SSM refers here and in the
following to standard solar models. In other words, our numerical
experiments disclose that a variation of the solar temperature in the
centre implies a definite variation in the entire inner radiative zone.

A consequence of this finding is that the different components of
the neutrino flux depend basically only on the central temperature,
and are almost independent of how that temperature is achieved.
This in turn implies that, when performing a $\chi^2$ analysis of the
experimental data compared to the prediction of non-standard solar
models, it is sufficient to parameterize these non-standard solar
models by the central temperature. In other words, varying
independently all the solar model parameters that influence the
temperature does not yield a better fit than just varying the central
temperature.

It is well known that it is not possible to get a good temperature fit
due to the ``discrepancy'' between the Kamiokande and Chlorine
results~\cite{Cast93a,Blud},
but the following questions are, nonetheless, interesting:

\noindent i)
how much does the fit improve if one excludes one of the
experimental results?

\noindent ii)
does this fit improve if one lowers  the
$p+{}^7\text{Be}\to \gamma+ {}^8\text{B}$ cross section,
as suggested  from the analysis of recent data on the Coulomb
dissociation of $^8\text{B}$~\cite{Langa,Moto}?

Another way to shift the nuclear fusion chain towards the $pp$-I
termination without altering the inner solar temperature
can be found in the realm of nuclear physics. In the light
of the new neutrino results, we discuss whether a hypothetical low
energy resonance in the $^3\text{He}+{}^3\text{He}$ reaction,
firstly advocated by
Fowler~\cite{Fowler}, analyzed in Ref.~\cite{Cast93a},
and presently investigated
experimentally at LNGS~\cite{Arpe91}, can reconcile theory and experiments.

Several new-generation experiments are being planned for the
detection of monochromatic solar neutrinos produced in electron
capture ($^7\text{Be}+e^- \to {}^7\text{Li}+\nu$)
and in the $pep$ ($p+e^- \to d+n$)
reactions~\cite{Arpe92,Alessa,Ragha}.
Furthermore, Bahcall~\cite{Bahcall93,Bahcall94a}
pointed out that thermal effects on
monochromatic neutrino lines can be used to infer inner solar
temperatures. In relation with the foregoing analysis, we discuss
what can be learnt from such future measurements about the
properties of neutrinos and of the Sun.

Concerning the organization of the paper, the solar-model-independent
constraints on neutrino fluxes are presented in Sec.~\ref{sec2}
and compared with the results of standard solar models in Sec.~\ref{sec3}.
Section~\ref{sec4} is devoted to the analysis of non-standard solar models
with lower temperature, which are compared with experimental data
in Sec.~\ref{sec5}.
In Sec.~\ref{sec6} we discuss the chances of a low energy resonance
in the $^3\text{He}+{}^3\text{He}$ channel, and in
Sec.~\ref{sec7} we remark the relevance of
future detection of the $pep$ and $^7\text{Be}$ neutrinos. Our conclusions are
summarized in the final Section.

\section{(Almost) solar model independent constraints on
          neutrino fluxes}
\label{sec2}

In this section we briefly update the constraints on neutrino fluxes
derived in Ref.~\cite{Cast94}, in the light of the recent reports from
Gallex and Sage~\cite{Gallex,SAGE}.
While we refer to Ref.~\cite{Cast94} for details, we recall here the main
points.

\noindent i)
For standard neutrinos and under the assumption of a nuclearly
powered Sun, the components $\Phi_i$ of the total neutrino flux arriving
onto the Earth are constrained by the equation of energy production
\begin{equation}
\label{energy}
    K = \sum_i \left(\frac{Q}{2} - \langle E \rangle_i\right)\,\Phi_i \quad ,
\end{equation}
where $K$ is the solar constant, $Q$ is the energy released in the fusion
reaction $4p+2e \to \alpha + 2\nu$
and $\langle E \rangle_i$ is the average neutrino energy of the $i$th flux.
In practice the relevant terms in Eq.~(\ref{energy}) are just those
corresponding to $\Phi_{pp+pep}$,
$\Phi_{\text{Be}}$, and
$\Phi_{\text{CNO}}$.

\noindent ii)
In order to calculate $\langle E \rangle_i$,
we take the ratio $\xi \equiv \Phi_{pep}/ \Phi_{pp+pep}$ from
the SSM ($\xi=2.38\times 10^{-3}$), and, similarly, the ratio
$\xi \equiv \Phi_{\text{N}}/ \Phi_{\text{CNO}}=0.54$.
Results are almost insensitive to these choices~\cite{Cast94}.

\noindent iii)
The signal $S_X$ of the $X$ experiment is represented as
\begin{equation}
\label{signal}
      S_X= \sum_i X_i \Phi_i \quad,
\end{equation}
where the weighting factors $X_i$ are cross sections for the $\nu$ detection
reaction averaged over the (emission) spectrum of the $i$-th
component of the neutrino flux (note that the $X_i$ are ordered
according to the neutrino energy), and are shown in Table~\ref{Xsect}.

\noindent iv)
We use the following experimental results, where systematic and
statistical errors have been added in quadrature. For the Gallium value,
we use the weighted average of the Gallex~\cite{Gallex}
and Sage~\cite{SAGE} results
\begin{equation}
\label{Gallium}
         S_{\text{Ga}}= (78 \pm 10)\text{ SNU} \quad .
\end{equation}
For the Chlorine experiment we use the average of the 1970-1992
runs~\cite{Davis}
\begin{mathletters}
\label{Boron}
\begin{equation}
\label{Cl}
         S_{\text{Cl}}= (2.32 \pm 0.26)\text{ SNU} \quad .
\end{equation}
Whereas the Kamiokande result reads
\begin{equation}
\label{Ka}
         S_{\text{B}}^{\text{Ka}}= (2.9 \pm 0.42)
            \times 10^6 \text{cm}^{-2} \text{s}^{-1} \quad .
\end{equation}
\end{mathletters}

\noindent v)
We take the Boron flux $\Phi_{\text{B}}$,
which enters in Eq.(~\ref{signal}), from
experiment. However, we can use {\em either} the Kamiokande result {\em or}
the Chlorine result (it is well known~\cite{Cast93a,Blud} that a choice
between the two
experiment is needed, otherwise one is forced to an unphysical value
$\Phi_{\text{Be}} \leq 0$).

We have thus four unknowns
$\Phi_{pp+pep}$,  $\Phi_{\text{Be}}$, $\Phi_{\text{CNO}}$, and
$\Phi_{\text{B}}$,
which are
constrained by the three equations~(\ref{energy}), (\ref{Gallium}),
and, alternatively, (\ref{Cl}) or (\ref{Ka}).

By exploiting the ordering properties of the $X_i$, as discussed in
Ref.~\cite{Cast94},
and by using the new experimental results, one derives severe constraints,
for standard neutrinos.
 As an example, by taking $\Phi_{\text{B}}$
from Kamioka, for each assumption about $\Phi_{pp+pep}$ one has the
minimum signal in Gallex if all other neutrinos are from Beryllium and
the maximum signal if all other neutrinos are from CNO.
By using similar procedures one finds
the bounds depicted in Figures.~\ref{fig2}, \ref{fig3},
and \ref{fig4}. By
conservatively using the Chlorine result to determine the Boron flux
(this choice is the less restrictive on the fluxes),  we find the
following bounds on the fluxes, in units of
$10^9 \text{cm}^{-2} \text{s}^{-1}$,
\begin{mathletters}
\label{bounds}
\begin{eqnarray}
64 \leq  \Phi_{pp+pep} &\leq& 65 \quad\quad\text{at 1 } \sigma\nonumber\\
      \Phi_{\text{Be}} &\leq& 0.7                             \nonumber\\
     \Phi_{\text{CNO}} &\leq& 0.6 \quad ,
\end{eqnarray}
and
\begin{eqnarray}
61 \leq  \Phi_{pp+pep} &\leq& 65 \quad\quad\text{at 3 } \sigma\nonumber\\
      \Phi_{\text{Be}} &\leq& 4.2  \nonumber\\
     \Phi_{\text{CNO}} &\leq& 3.6 \quad .
\end{eqnarray}
\end{mathletters}
In summary, the Gallium result together with the luminosity
constraint implies that almost all neutrinos, if standard, come from
the $pp$-I termination. The bounds of Eqs.~(\ref{bounds})
are very strict  since
even a small flux of other (and more energetic) than the $pp$ neutrinos
gives
an appreciable contribution to the Gallium signal. This is why an
experimental result with 10\% accuracy can fix the
$\Phi_{pp+pep}$ at the level of about 2\%.

We note that the bounds have become much more stringent than
those reported in Ref.~\cite{Cast94}, because both the central value and the
error of the Gallium result have decreased, so that now the
experimental result is even closer to the minimal signal which is
obtained when all neutrinos come from the $pp$-I termination
($\Phi_{pp+pep}= 65\times 10^9 \text{cm}^{-2} \text{s}^{-1}$).

Concerning the assumptions leading to Eqs.~(\ref{bounds}),
we remark that the main hypothesis is that the present Sun is nuclearly
powered, see Eq.(~\ref{energy}),
whereas the values chosen for $\xi$ and $\eta$ are unessential (see
again Ref.~\cite{Cast94}).
\section{Standard solar models and experimental data}
\label{sec3}
The relevance of the bounds derived in the previous section can be
best illustrated by comparing them with the results of standard solar
model computations. For a few representative calculations we
present the main input parameters of these
models in Table~\ref{Inpar}, and the resulting neutrino
fluxes in Table~\ref{SMflux}.

Let us remark that we can now compare not only the total signals
predicted by the theory and measured by experiments, but also
several individual fluxes, as shown in Table~\ref{SMflux}.
In particular, we find that the upper limit for $\Phi_{\text{Be}}$,
implied by the
experiment at the $1 \sigma$ level, is 7 times smaller than
$\Phi_{\text{Be}}^{\text{SSM}}$, whereas,
at the same level of accuracy, the suppression of
$\Phi_{\text{B}}$ is about a
factor of two respect to the SSM (in Table~\ref{SMflux}
the experimental upper
bound on $\Phi_{\text{B}}$ is obtained from the less constraining result,
i.e. the Kamiokande value). A suppression of $\Phi_{\text{Be}}$
stronger than $\Phi_{\text{B}}$ was
already implied by the comparison between Kamiokande and
Chlorine results, while we derived it using essentially only the
Gallium experiments.

In addition, we remind that the theoretical calculation for
$\Phi_{\text{B}}$ is the
most questionable of the flux calculations, due to the well known
uncertainties. In our opinion, the discrepancy between theory and
experiment for the $^7{\text{Be}}$ flux
is much more serious than the one for the $^8\text{B}$ flux.
In other words, it seems to us that the solar neutrino problem is now
at the level of the branching between the $pp$-I and $pp$-II
terminations.

In order to reconcile the theoretical and experimental determination
of $\Phi_{\text{B}}$, one needs that the ratio between the two rates for the
$^3\text{He}+{}^4\text{He}$  and the $^3\text{He}+{}^3\text{He}$ reactions,
\begin{equation}
  R = \frac{ \langle \lambda_{34} \rangle}{ \langle \lambda_{33} \rangle}
   \quad ,
\end{equation}
is drastically altered from $R^{\text{SSM}}=0.16$ to something about
$R=0.02$
(here and in the following, $\lambda_{ij}$ is the rate for the
collision between
nuclei with mass number $i$ and $j$, $m_{ij}$ being the reduced mass).

The investigation of non-standard solar models where $R$ is strongly
reduced will be the subject of the next sections. It is worth
remarking however that a reduction of $\Phi_{\text{Be}}$ to bring it in
the experimentally acceptable range
generally implies also a comparable, or even larger, reduction of
$\Phi_{\text{B}}$,
which then becomes too small with respect to the experimental
value.

\section{Non-standard solar models with low central temperature}
\label{sec4}

Clearly  the $pp$ chain can be shifted towards the $pp$-I termination by
lowering the inner temperature $T$, since the tunnelling probability is
more reduced for the heavier nuclei:
\begin{equation}
 \log \left(
   \frac{ \langle \lambda_{34} \rangle}{ \langle \lambda_{33} \rangle}
       \right) \propto
  \frac{m_{33}^{1/3} - m_{34}^{1/3}}{(KT)^{1/3}} \quad .
\end{equation}
In order to reduce the inner temperatures one may attempt several
manipulations~\cite{Cast94}:

\noindent i)
reduce the metal fraction Z/X,

\noindent ii)
reduce (by an overall multiplicative factor) the opacity tables,

\noindent iii)
increase the astrophysical factor $S_{pp}$ of the
$p+p\to d + e + \nu $ reaction,

\noindent iv)
reduce the Sun age.

Clearly i) and ii) work in the direction of getting a more transparent
Sun, which implies a lower temperature gradient, a larger energy
production region and
consequently smaller inner temperatures. When $S_{pp}$ is increased
nuclear fusion gets easier, and the fixed luminosity is obtained with a
reduced temperature. A younger Sun is another way to get a
Sun cooler in its interior, since the central H-abundance is increased
and, again, nuclear fusion gets easier.

On the other hand, we remark that variations of  the other
astrophysical $S$-factors, $S_{33}$, $S_{34}$ and/or $S_{17}$,
affect very weakly the
inner solar temperature. This is physically clear, since the energy
production mechanism is untouched~\cite{Cast93a}.

We have computed several solar models by varying the parameters
well beyond the uncertainties of the standard solar model
(see Table~\ref{thomo}),
i.e. we have really built non-standard solar models.

An important feature of all these models is the homology of the inner
temperature profiles
\begin{equation}
\label{homo}
  T(m)= k T^{\text{SSM}}(m) \quad ,
\end{equation}
where $m=M/M_0$ is a mass coordinate, and the factor $k$ depends on the
parameter which is varied but does not depend on $m$.

We have verified that Eq.~(\ref{homo}) holds with an accuracy better than 1\%
in all the internal radiative zone ($M/M_0<0.97$ or
$R/R_0<0.7$) for all the
models we consider, but for huge (and really unbelievable)
variations of the solar age, see Fig.~\ref{fig5} and Table~\ref{thomo}.
It is worth
remarking that $T(m)/T^{\text{SSM}}(m)$ is constant through a region where
$T(m)$ change by a factor five, see Fig.~\ref{fig6}.

By looking at the numerical results, one finds - as expected - that, as long
as the Sun
age is kept fixed, the models have similar distributions of $^{4}$He and of
the energy production per unit mass, which as well known, is strongly
related with temperature and $^{4}$He density.
On the other hand, when the Sun age is varied, the $^{4}$He content
also changes strongly, and the homology relation for the temperature
is fading away.  The important point is that for each model the
temperature profile is essentially specified
by a scale factor, which can be taken as the central temperature $T_c$.

On these grounds one derives general predictions for the behaviour
of the neutrino fluxes $\Phi_i$. They are crucially dependent (through the
Gamow factors) on the values of the temperature in the production
regions $T_i$, and, as usual, can be locally approximated by power laws:
\begin{equation}
\label{fplaw}
 \Phi_i= c_i \, T_i^{\beta_i} \quad.
\end{equation}
The homology relationship implies
$T_i=(T_c/T_c^{\text{SSM}}) T_i^{\text{SSM}}$ and,
consequently,
\begin{equation}
\label{ftdep}
  \Phi_i = \Phi_i^{\text{SSM}}\,
            \left(\frac{T_c}{T_c^{\text{SSM}}}\right)^{\beta_i}  \quad.
\end{equation}
This means that each flux is mainly determined by the central
temperature, almost independently on the way the temperature
variation was obtained, an occurrence which is
clearly confirmed by Fig.~\ref{fig7} for the components of the neutrino
flux which give the main contributions
($\Phi_{pp}$, $\Phi_{\text{Be}}$, and $\Phi_{\text{B}}$)
to the experimental signals.

The situation is shown in more details in
Table~\ref{betas} where we present
the numerically calculated values  of the $\beta_i$ coefficients. One sees that
$\beta_{pp}$, $\beta_{\text{Be}}$, and $\beta_{\text{B}}$
are approximately independent on the parameter
which is varied. This is not true for
$\Phi_{\text{N}}$, $\Phi_{\text{O}}$, and $\Phi_{pep}$.
Actually,
when writing Eq.~(\ref{fplaw}) we neglected the flux dependence on the
densities of the parent nuclei which generate solar neutrinos. These
densities can change when some of the input parameters are
varied. For example, $\Phi_{\text{N}}$ and $\Phi_{\text{O}}$
look very sensible to variations of
$Z/X$, since in this case, in addition to the temperature variation, the
change of metallicity also influences the effectiveness of the CN cycle.
However, this effect is negligible when estimating total experimental
signals.

Analytical approximations to the
numerical values of the $\beta_i$ can be found
by considering  the dependence on temperature of the
Gamow factors for the relevant nuclear reactions~\cite{Cast93b}.
We would
like to comment here just on the temperature dependence of  the
ratio  $\Phi_{\text{B}} / \Phi_{\text{Be}}$:
\begin{equation}
 \frac{\Phi_{\text{B}}}{ \Phi_{\text{Be}}} =
  \frac{n_p\langle \sigma_V \rangle_{17}}
       {n_e\langle \sigma_{V_e} \rangle_{\text{capt}}}
   \propto \frac{n_p}{n_e} \frac{T^{\gamma_{17}}}
                                {T^{\gamma_{\text{capt}}}} \quad ,
\end{equation}
where $\gamma_{\text{capt}}=-1/2$, and $\gamma_{17}=-2/3+E_{17}/KT$
($E_{17}$ is
the Gamow peak for the
$p+{}^7\text{Be}\to \gamma+ {}^8\text{B}$
reaction)~\cite{Rolfs}. Assuming $n_p/n_e$ to be constant,
and evaluating  $E_{17}/KT$ at $T_c^{\text{SSM}}$, we get
\begin{equation}
\frac{\Phi_{\text{B}} }{ \Phi_{\text{Be}}} \propto T_c^{13.5} \quad .
\end{equation}
This value is in good agreement with the one reported in Table~\ref{betas}
for a $S_{pp}$  variation; the agreement is less good with the values
obtained by varying the other parameters (in this case $n_p/n_e$ is
clearly not conserved).

Therefore, as long as the temperature profile is {\em unchanged}, lowering the
temperature immediately implies that Boron neutrinos are suppressed
much more strongly than Beryllium neutrinos, since the penetrability
factor for the
$p+{}^7\text{Be}\to \gamma+ {}^8\text{B}$
reaction is diminished.
\section{The central solar temperature and the experimental results.}
\label{sec5}
{}From the argument just presented, it is clear that a central
temperature reduction cannot work; nevertheless, let us perform a
$\chi^2(T_c)$ analysis to see quantitatively what happens. We define:
\begin{equation}
\chi^2(T_c)=\sum_{XY} ( S^{\text{ex}}_X -
                        S^{\text{th}}_X ) V^{-1}_{XY}
                      ( S^{\text{ex}}_Y -
                        S^{\text{th}}_Y ) \quad ,
\end{equation}
where the symbols have the following meaning.

\noindent i)
The experimental signals $S^{\text{ex}}_X$ ($X=$  Gallium, Chlorine and
Kamiokande) are the ones reported in Eqs.~(\ref{Gallium}) and (\ref{Boron}).

\noindent ii)
The theoretical signals $S^{\text{th}}_X(T_{c})$ are calculated according
to the formula
\begin{equation}
S^{\text{th}}_X = \sum_{i\neq pp} X_i \,\Phi_i^{\text{SSM}}
                  \left(\frac{T_c}{T_c^{\text{SSM}}}\right)^{\beta_i}
               + X_{pp} \, \Phi_i^{pp} \quad ,
\end{equation}
where we take the $\beta$ coefficients corresponding to the $S_{pp}$
variations
(second column of Table~\ref{betas}), and we use the
CDF94 standard solar
model results, see Table~\ref{SMflux}. Note, in particular, that
$\Phi_{\text{B}}^{\text{SSM}}$ has been
calculated by using $S_{17}=22.4$ eV barn. In order to achieve a better
accuracy, $\Phi_{pp}$ is calculated directly through the Eq.~(\ref{energy}).

\noindent iii)
The error matrix $V_{XY}$ takes into account both the experimental
and the theoretical uncertainties. The theoretical uncertainties are
due to the neutrino cross sections $X_i$, and to the solar model
parameters that are not related to the free parameter $T_c$, i.e.
$S_{33}$, $S_{34}$, and $S_{17}$.
The diagonal entries, $V_{XX}$, are the sum of the experimental
variance $\sigma_X^2$, plus the the squares of the errors due to the
cross sections $\sum_i(\Delta_X^{i})^2$ ($\Delta_X^{i}$ is the error of the
detection cross section for the $X$ experiment averaged over the $i$-th flux),
plus the squares of the
errors due to the input parameters $S_{33}$, $S_{34}$, and  $S_{17}$, i.e.
$\sum_P(\Delta_X^{P})^2$ ($P= S_{33},S_{34}, S_{17}$).
The off-diagonal entries have contributions only from these last errors:
$V_{XY}= \sum_P\Delta_X^{P}\Delta_Y^{P}$.  The errors $\Delta$
are calculated by linear propagation. Therefore, if we call
$\delta_X^{i}$ the error on the cross section $X_i$,
$\Delta_X^{i} =  \Phi_i^{\text{SSM}}
                  \left(\frac{T_c}{T_c^{\text{SSM}}}\right)^{\beta_i}
                    \delta_X^{i}$,
while, if $\delta^{P}$ is the error on the
parameter $P$,
$\Delta_X^{P} = \left( \partial S^{\text{th}}_X / \partial P \right)
\delta^{P}$. The the partial derivative of the neutrino fluxes respect
to these parameters are estimated by using power-laws which we
have been determined from numerical experiments, and which are very
similar to those of Table~7.2 in Ref.~\cite{Bahcall89}.
The values we use for the
uncertainties of the SSM parameters, $\delta^{P}$, are given in
Table~\ref{SMpar},
while the errors on the cross sections, $\delta^{i}_X$,
can be found in Table~\ref{Xsect}. The
use of the error matrix is necessary to avoid that an apparently good
fit be achieved in an unphysical way, e.g. we cannot use the
uncertainty of the Boron flux to strongly reduce its contribution to
the Davis experiment, and, at the same time, have a smaller
reduction in the Kamiokande experiment.

The results shown in Fig.~\ref{fig8}(a) deserve a few comments.

\noindent i)
The best fit to the three experimental signals yields a
$\chi^2_{\text{min}}$[Cl+Ga+Ka]=
18.5 that, for two degrees of freedom, is excluded at the 99.99\%
level (here we have treated
systematic and statistical errors on equal footing);
we thus confirm the results of Ref.~\cite{Blud93}.
This is partly due to
the well known ``inconsistency'' between Kamiokande and Chlorine.

\noindent ii)
Even if we only consider Gallium and Kamiokande the fit is still
poor, yielding a $\chi^2_{\text{min}}$[Ga+Ka]= 11,
that for one degree of freedom is
excluded at the 99.9\% level. The reason is that if one tries to reduce
$\Phi_{\text{Be}}$ in accordance with Gallium data,
then $\Phi_{\text{B}}$ becomes too small in
comparison with the Kamiokande result. On the other hand, if one
considers just Gallium and Chlorine results the situation is better
($\chi^2_{\text{min}}$[Cl+Ga]= 5, which has a 2.5\% probability),
due to the fact that
the smaller Boron (and Beryllium) signal implied by the Chlorine
experiment can be more easily adjusted to the Gallium result.

\noindent iii)
{}From the above discussion it is clear that if one lowers the
$p+{}^7\text{Be}\to \gamma+ {}^8\text{B}$
cross section, the situation gets even worse, see Fig.~\ref{fig8}(b).
In other words, a reduction of S17 does not solve  the solar
neutrino problem.

\noindent iv)
Considering the Chlorine data  corresponding  (approximately) to
the same data taking period as the other experiments
($S_{\text{Cl}}^{88-92}=
2.76 \pm 0.31 \text{ SNU}$~\cite{Davis}) the
situation is only slightly changed:
$\chi^2_{\text{min}}$[Cl+Ga+Ka]= 15
that, for two degrees of freedom, is
excluded at the 99.94\% level; $\chi^2_{\text{min}}$[Ga+Ka]=
11, that for one degree of
freedom is excluded at the 99.9\% level; and
$\chi^2_{\text{min}}$[Cl+Ga]=6, which
has a 2.4\% probability.

\noindent v)
For the uncertainties of Table~\ref{SMpar},
the effect  of the error correlation is not large:
for instance, if we use
uncorrelated errors
$\chi^2_{\text{min}}$[Cl+Ga+Ka]= 16
instead of 18.5.
The real importance of error correlation
becomes evident if we try to resolve the
discrepancy by increasing the errors. For example, doubling the
uncertainties reduces the uncorrelated
$\chi^2_{\text{min}}$ to 14, while the
correlated one practically does not change.

\noindent vi)
The situation does not significantly change
when considering models where one of the other parameters (opacity table,
Z/X, age) are varied instead of $S_{pp}$,
as it is shown by  Fig.~\ref{fig9}. Slightly better fits are
obtained by varying Z/X or the age than $S_{pp}$ or the opacities,
but the resulting
$\chi^2_{\text{min}}$[Cl+Ga+Ka]= 16.5
is still excluded at the 99.97\% level.

\noindent vii)
If one insists on a low temperature solution, the
best fit is for $T_c/T_c^{\text{SSM}}\approx 0.94$,
i.e.  $T_c =1.46 \times 10^7 \,{}^oK$. The price to pay
for this 6\% temperature reduction is very high in terms of the input
parameters which are being varied, see Table~\ref{SMpar}.
Huge variations of
the parameters are required, and, furthermore, in many cases  the
values used are at the border of what can be tolerated by our stellar
evolution code: for example, we are not able to produce a Sun with
$T_c/T_c^{\text{SSM}}<0.94$  by lowering the opacity  or the age.
\section{A low energy resonance in the
         $^{\bbox{3}}\protect\text{He} + {}^{\bbox{3}}\protect\text{He}$
          channel?}
\label{sec6}
As mentioned in the introduction, the other way to enhance the $pp$-I
termination is to play with the $^3\text{He}$ nuclear cross sections.
As it was shown in Ref.~\cite{Cast93a}, if the astrophysical S-factors
are varied by a constant (through the star) quantity:
\begin{mathletters}
\label{s34scale}
\begin{equation}
           \Phi_{i}=
           \Phi_{i}^{\text{SSM}} \, \theta
\end{equation}
where
\begin{equation}
              \theta = \frac{S_{34}}{S_{34}^{\text{SSM}}}
                      \sqrt{ \frac{S_{33}^{\text{SSM}}}{S_{33}} }
\quad\quad \text{ and } \quad i = \text{B, Be}  \quad .
\end{equation}
\end{mathletters}
Numerical experiments confirm the approximate validity of
Eqs.~(\ref{s34scale})
giving
$\Phi_{\text{B}\, ,\text{Be} } =
           \Phi_{\text{B}\, ,\text{Be}}^{\text{SSM}}\, \theta^{0.9}$.
Note that the changes of $\Phi_{\text{B}}$ and
$\Phi_{\text{Be}}$ are proportional.

For variations of $S_{33}$ and $S_{34}$ the solar temperature is essentially
unaffected, and, consequently, all the fluxes other than B and Be are also
unaffected. Only the $pp+pep$
neutrino flux slightly changes, in order to fulfill the
luminosity condition, Eq.~(\ref{energy}), i.e.
\begin{equation}
 \Phi_{pp+pep} = \Phi_{pp+pep}^{\text{SSM}} +
                 \Phi_{\text{Be}}^{\text{SSM}} - \Phi_{\text{Be}}
\end{equation}
In order to reduce the Beryllium flux by a factor -- say -- three with
respect to the SSM value, $S_{33}$ ($S_{34}$) has to be nine times (one third)
the value used in the standard solar model calculations. Clearly, what
matters are the values of the astrophysical factors at the energies
relevant in the Sun, i.e. at the position of the Gamow peak  for the
He + He reactions near the solar center, $E_{\text{G}}\approx 20 \text{ keV}$.

We recall that the astrophysical factors used in the calculations are
obtained by extrapolating experimental data taken at higher energies
(see Ref.~\cite{Rolfs} for a review).  Thus a very low energy resonance in
the $^3\text{He}+{}^3\text{He}$ reactions could be effective in reducing
$ \Phi_{\text{Be}}$ and $\Phi_{\text{B}} $, and
could have escaped to experimental detection. This possibility, first
advanced in Ref.~\cite{Fowler}, cannot be completely dismissed,
(see the discussion
in Refs.~\cite{Cast93a,Rolfs})
and it is presently being investigated in the underground
nuclear physics experiment LUNA at Laboratori Nazionali del Gran
Sasso~\cite{Arpe91}.

For a resonance at energy $E_r$ and with strength $\omega\gamma$,  \
equations~(\ref{s34scale}) become:
\begin{mathletters}
\label{resona}
\begin{equation}
           \Phi_{i}=
           \Phi_{i}^{\text{SSM}} \,
           \sqrt{\frac{1}{1+x_i} }
\quad\quad \text{   } \quad i = \text{B, Be}  \quad ,
\end{equation}
where
\begin{equation}
              x_i = \frac{\omega\gamma}{W}
                    \exp [3A(KT_i)^{-1/3}-E_r/(KT_i)]  \quad ,
\end{equation}
\end{mathletters}
\noindent and $T_i$ are the temperatures at the peak of the $\nu_{\text{Be}}$
and $\nu_{\text{B}}$ production
($T_{\text{Be}}=1.45\times 10^7\,{}^o\text{K}$,
$T_{\text{B}}=1.5\times 10^7\,{}^o\text{K}$), $K$ is the Boltzmann constant,
and the other constants, defined in
Ref.~\cite{Cast93a}, are  $W=20.4\text{ keV}$ and $A=1.804 \text{ MeV}^{1/3}$.

Let us remark that the resonance can work differently in different
regions of the Sun, in relationship with the kinetic energies of the
colliding particles. A low energy resonance is more efficient in the
outer zone of energy production, and consequently $\Phi_{\text{Be}}$ can be
suppressed more than $\Phi_{\text{B}}$.
The opposite occurs for higher energy
resonances, the turning point being $E_r\approx E_{\text{G}}$, see
Ref.~\cite{Cast93a} for details.

We have performed a $\chi^2$ analysis as a function of the resonance
strength $\omega\gamma$ for several values of the resonance energy
$E_r$, with a
procedure quite  similar to that used in the previous section.

The errors on the calculated signals arise from the neutrino
interaction cross sections,  from  $S_{17}$, and from all those quantities
which influence the estimated central temperature of the Sun ($S_{pp}$,
Z/X, opacity and age), but not from $S_{33}$ and $S_{34}$ that influence
fluxes according to Eq.~(\ref{s34scale}),
and correspond to our free parameter.
Again, the derivative of the neutrino fluxes with respect to these
parameters, necessary to calculate the error matrix by linear
propagation, are estimated by using power-laws  very similar to
those of Table 7.2 in Ref.~\cite{Bahcall89}.

The uncertainties we use are shown in Tables~\ref{SMpar} and \ref{Xsect}.
We note that uncertainties on
the absorption cross sections, the metallicity Z/X and the opacity are
the most important for estimating the errors on the signal.
For the opacity we followed
Ref.~\cite{Bahcall92} and took
``the characteristic difference between the solar
interior opacity calculated with Livermore and with Los Alamos
opacity code'', which may or may not be a fair estimate of the
uncertainty, but we could not find a better prescription. However,
as we shall see, the
correlation among the errors is such that $\chi^2_{\text{min}}$
does not change even
if we double the uncertainties on Z/X and on the opacity.

The results are presented in Fig.~\ref{fig10}. The situation looks slightly
better than in the low temperature models since the
$\Phi_{\text{Be}}$ reduction
does not imply an even stronger $\Phi_{\text{B}}$
reduction. However, the best $\chi^2_{\text{min}} = 14$,
obtained for $E_r = 0$, is still excluded at the 99.9\% level.
The $\chi^2_{\text{min}}$ slightly increases with $E_r$
because of the tuning of the
Beryllium/Boron suppression.

The best fit strength as a function of $E_r$ is shown in Fig.~\ref{fig11},
together
with existing experimental upper bound. We expect that LUNA
experiment, presently performed at LNGS~\cite{Arpe91}, will have a sensitivity
better by about a factor 100, as compared with previous experiments,
mainly due to the cosmic ray shielding
in the underground laboratory, so that the search should be able
to detect/exclude  such a resonance down to extremely low values of
$E_r$.

The use of the properly correlated errors on the fluxes  is even more
important when studying the effect of the hypothetical resonance
than when we changed the temperature. The $\chi^2_{\text{min}}$
would be 10
instead of 14, had we used uncorrelated errors. Moreover, doubling
the errors would yield a $\chi^2_{\text{min}}$
of almost 6, while the correlated one remains 14.
The intuitive explanation
of how the uncorrelated fit works is the following.
The Chlorine and Kamiokande results require different suppressions of the
neutrino fluxes. The fit finds the best compromise between the two
experiments by adjusting the resonance strength.
Then, the uncertainty on the temperature is used
to further deplete
$\Phi_{\text{Be}}$ and, at the same time, to increase $\Phi_{\text{B}}$,
which is clearly unphysical. The correlated fit correctly uses the
uncertainty on the temperature either to increase or to decrease both fluxes
at the same time: either
option is useless, once we get the best compromise for the common
reduction of the two fluxes, no matter how much we are allowed to change the
temperature.

Combining the two mechanisms, i.e. a resonance in a low
temperature model, does not work either, since again, once the best compromise
suppression of the $^7\text{Be}$ and $^8\text{B}$ fluxes is achieved by one of
the two mechanisms, the other cannot do much more.
\section{The detection of \lowercase{$\bbox{pep}$} and
$^{\bbox{7}}$B\lowercase{e} neutrinos}
\label{sec7}
New generation  experiments are being planned for the detection of
monochromatic solar neutrinos produced in electron capture
($^7\text{Be} + e^- \to {}^7\text{Li} + \nu$)
and in the $pep$ ($ p + e^- + p \to  d + \nu$)
reactions~\cite{Arpe92,Alessa,Ragha}.
Furthermore, Bahcall~\cite{Bahcall93,Bahcall94a}
pointed out that, from the measurement of the average energy difference
between neutrinos emitted in solar and laboratory decay, one can infer
the temperature of the production zone.
In this section we discuss what can be learnt from such future
measurements about the properties of neutrinos and of the Sun.

Concerning the intensity of the $^7\text{Be}$ line,
we recall the bounds of Eqs.~(\ref{bounds}):
at $1 \sigma$ ($3 \sigma$)
the neutrino flux has to be smaller than
$0.7\times 10^9\text{cm}^{-2}\text{s}^{-1}$
($4.0\times 10^9\text{cm}^{-2}\text{s}^{-1}$),
otherwise neutrinos are non-standard. We recall however
that a low
$\Phi_{\text{Be}}$ is also typical of the MSW solution, see Fig.~\ref{fig12}.

The $pep$ neutrinos are a good indicator of $\Phi_{pp}$, since the ratio
$\Phi_{pep}/\Phi_{pp}$ is rather stable.
In Fig.~\ref{fig12} we see that standard neutrinos
correspond to $\Phi_{pep}$ in the range $(1 \div 2)\times 10^8
\text{cm}^{-2}\text{s}^{-1}$, whereas the MSW
solution requires $\Phi_{pep} \leq 3 \times 10^7 \text{cm}^{-2}\text{s}^{-1}$.
Thus, a measurement of  the
$pep$-line intensity will be crucial for deciding about neutrino
properties.

The possibility of measuring inner solar temperatures through thermal
effects
on monochromatic neutrino lines looks to us extremely
fascinating (although remote). In this respect the homology
relationship, Eq.~(\ref{homo}),
is particularly interesting, see Fig.~\ref{fig13}.

If homology holds,  a measurement of the solar temperature  in the --
say -- $^7\text{Be}$ production zone gives the value of $T_c$.
On the other hand,
the homology relation itself is testable -- in principle -- by comparing
the temperatures at two different places, as can be done by looking
at the shapes of both the
$\nu_{\text{Be}}$ and $\nu_{pep}$ lines. We remark that this
would be a test of the mechanism for energy transport through the
inner Sun.
\section{Conclusions}
\label{conclusions}
i)
If neutrinos are standard, the present solar neutrino experiments
already impose severe constraints on the individual components of
the total neutrino flux. These constraints, at the $1 \sigma$ level, are:
\begin{eqnarray}
\label{bound1s}
\Phi_{\text{Be}}  &\leq& 0.7 \times 10^9 \text{cm}^{-2} \text{s}^{-1}
\nonumber\\
\Phi_{\text{CNO}} &\leq& 0.6 \times 10^9 \text{cm}^{-2} \text{s}^{-1}
\nonumber\\
64 \times 10^9 \text{cm}^{-2} \text{s}^{-1} \leq
       \Phi_{pep} &\leq& 65 \times 10^9 \text{cm}^{-2} \text{s}^{-1}
\end{eqnarray}
The constraint on Beryllium neutrinos is in strong disagreement with
the results of any standard solar model calculation, see Table~\ref{SMflux}.
The solar neutrino problem is now at the Beryllium production level: the
experimental data demand a strong shift towards the $pp$-I
termination,  and the problem is not restricted anymore to the rare
$pp$-III ($^8$B) termination.

ii)
Solar models with low inner temperatures
show  temperature profiles $T(m)$
homologous to that of the Standard Solar Model: $T(m)=k T^{\text{SSM}}(m)$.
As a consequence,  the  main components of the neutrino flux depend
essentially on the central solar temperature $T_c$ (see
Table~\ref{thomo}), and
the experimental signals can be parameterized in terms of $T_c$. As
already known, there is no value of $T_c$ which can account for all the
available experimental results
($\chi^2_{\text{min}}(T_c)\approx 16$). In addition, we find
that the fit is poor even  considering  just Gallium and Kamiokande
results ($\chi^2_{\text{min}}(T_c)\approx 11$).
Furthermore, lowering the cross section for
$p+{}^7\text{Be} \to \gamma + {}^8\text{B}$ makes things worse.

iii)
Alternatively, the shift of the nuclear fusion chain towards the $pp$-I
termination could be induced by a hypothetical low energy
resonance in the $^3\text{He} + {}^3\text{He}$ reaction.
This mechanism gives a somehow
better but still poor fit to the combined experimental data
($\chi^2_{\text{min}}(T_c)\approx 14$).
Its possible relevance to the solar neutrino problem
will be elucidated in an underground nuclear physics experiment,
presently performed at LNGS.

iv)
Concerning future experiments, the measurement of the $^7\text{Be}$ and,
particularly, of the $pep$-line intensities will be crucial for
discriminating non-standard solar models from non-standard
neutrinos, in relation with the bounds in Eq~(\ref{bound1s}).
Furthermore, the
homology relation itself can be tested, in principle, in experiments
aimed at the measurement of  inner solar temperatures by looking at
thermal effects on
the  $pep$ and Be line shapes.
This would provide
a clear test about the mechanism of energy transport in the solar
interior.

In conclusion,  we feel that recent Gallium results, taken at their face
value, strongly point towards non-standard neutrinos. Of course we
are anxiously waiting for the calibration of Gallex and Sage, and for
future experiments.

\acknowledgments
One of us (G.~F.) acknowledges useful discussions with V.~Berezinsky.
\begin{table}
\caption[aaaa]{
The main parameters $P$ of solar models  and their estimated
relative uncertainties at $1 \sigma$ level,
$(\delta^P/P)^{\protect\text{SSM}}$ (here as in the text
$\delta^P\equiv \delta P$).
All values are as in Ref.~\protect\cite{Bahcall92}, apart
for $S_{pp}$,
which is taken from the  more recent Ref.~\protect\cite{Bahcall94b}.
Concerning
solar age we refer to common wisdom,
see Ref.~\protect\cite{Turck93a}. In the last column
we show, for the first four parameters, the values of
$\zeta \equiv P/ P^{\protect\text{SSM}}$
needed to account for
$T_c/T_c^{\protect\text{SSM}} =0.94$, when each  input parameter is
varied separately.  In the same column we also show, for $S_{33}$
and $S_{34}$,
the values needed to account for
$\Phi_{\protect\text{Be}} = 0.3 \,
 \Phi_{\protect\text{Be}}^{\protect\text{SSM}} $
(again when each input parameter is varied separately).
   \label{SMpar}
               }
\begin{tabular}{lr@{}ld}
  $P$       &\multicolumn{2}{c}{
    $\left(\frac{\delta^P}{P}\right)^{\protect\text{SSM}}$}
                     &\multicolumn{1}{c}{$\zeta
                    = \frac{P}{P^{\protect\text{SSM}}}$}\\
\tableline
$S_{pp}$    &  1 &  \%   & 1.7   \\
opacity     &  2 &.5\%   & 0.63  \\
Z/X         &  6 &  \%   & 0.30  \\
age         &  3 &  \%   & 0.23  \\
$S_{33}$    &  6 &  \%   &11.0   \\
$S_{34}$    &  3 &  \%   & 0.3   \\
$S_{17}$    &  9 &  \%   & \multicolumn{1}{c}{--} \\
\end{tabular}
\end{table}
\begin{table}
\caption[bbbb]{
   For the $i$-th component of the neutrino flux we show the
average neutrino energy $\langle E \rangle$ and
the averaged neutrino capture cross
sections $X_i$
(1 SNU cm$^2$ s = $10^{-36}$ cm$^2$) for Chlorine (Cl) and Gallium (Ga),
with errors at $1 \sigma$ level.
All data are from Ref.~\protect\cite{Bahcall89},
but for the Cl cross section average over the $^8$B neutrino flux, which is
taken from Ref.~\protect\cite{Garcia}.
When averaging the $pp$ and $pep$ components we use the relative
weights of our SSM (CDF94), see Table~\protect\ref{Inpar};
similarly for $^{13}$N and $^{15}$O.
   \label{Xsect}
               }
\begin{tabular}{lr@{}lr@{}l@{}lr@{}l@{}l}
         &\multicolumn{2}{c}{$\langle E \rangle$}
                    &\multicolumn{3}{c}{Cl}
                                                 &\multicolumn{3}{c}{Ga} \\
         &\multicolumn{2}{c}{[MeV]}
                    &\multicolumn{3}{c}{[10$^{-9}$SNU cm$^2$s]}
                         &\multicolumn{3}{c}{[10$^{-9}$SNU cm$^2$s]} \\
\tableline
$pp$     & 0. & 265 &    0. &    &               &    1. & 18 & $(1\pm 0.02)$\\
$pep$    & 1. & 442 &    1. & 6  & $(1\pm 0.02)$ &   21. & 5  & $(1\pm 0.07)$\\
$pp+pep$ & 0. & 268 &       &    &               &    1. & 23 & $(1\pm 0.02)$\\
$^7$Be   & 0. & 814 &    0. & 24 & $(1\pm 0.02)$ &    7. & 32 & $(1\pm 0.03)$\\
$^{13}$N & 0. & 707 &    0. & 17 & $(1\pm 0.02)$ &    6. & 18 & $(1\pm 0.03)$\\
$^{15}$O & 0. & 996 &    0. & 68 & $(1\pm 0.02)$ &   11. & 6  & $(1\pm 0.06)$\\
CNO      &    &     &       &    &               &       &    &              \\
($^{13}$N + $^{15}$O)
         & 0. & 840 &    0. & 40 & $(1\pm 0.02)$ &    8. & 67 & $(1\pm 0.05)$\\
$^8$B    & 6. & 71  & 1090. &    & $(1\pm 0.01)$ & 2430. &    & $(1\pm 0.25)$\\
\end{tabular}
\end{table}
\begin{table}
\caption[cccc]{
   Physical input parameters of several Standard Solar
Models. We show the solar mass $M_0 [10^{33}\protect\text{gr}]$,
the solar radius $R_0 [10^{10}\protect\text{cm}]$,
the solar luminosity $L_0 [10^{33}\protect\text{erg/s}]$,
the solar age $[10^{9}\protect\text{yr}]$,
the metal to hydrogen mass fraction Z/X, the zero energy
astrophysical $S$-factors
[MeV barn] and their derivatives with respect to energies $S'$ [barn].
BP is ``the best model with diffusion'' of Ref.~\protect\cite{Bahcall92};
TCL is the ``IS Cpp Recent CNO model'' of Ref.~\protect\cite{Turck93b};
CDF94 is our updated standard solar model, with Livermore opacity
table~\protect\cite{Iglesias},
chemical composition following Grevesse 1991
``low iron''~\protect\cite{Greve}, and
without diffusion.
   \label{Inpar}
               }
\begin{tabular}{lr@{}l@{}lr@{}l@{}lr@{}l@{}l}
physical                                    &\multicolumn{3}{c}{BP}
             &\multicolumn{3}{c}{TCL}       &\multicolumn{3}{c}{CDF94} \\
quantities   & & & \\
\tableline
$M_0$                                       &  1. & 989&
             &  1. & 989&                   &  1. & 989&                  \\
$R_0$                                       &  6. & 96 &
             &  6. & 96 &                   &  6. & 96 &                  \\
$L_0$                                       &  3. & 86 &
             &  3. & 85 &                   &  3. & 83 &                  \\
Age                                         &  4. & 6  &
             &  4. & 5  &                   &  4. & 6  &                  \\
Z/X                                         &  2. & 67 & $\times 10^{-2}$
             &  2. & 43 & $\times 10^{-2}$  &  2. & 67 & $\times 10^{-2}$ \\
\tableline
$S(0)_{pp}$                                 &  4. & 00 & $\times 10^{-25}$
             &  4. & 00 & $\times 10^{-25}$ &  3. & 89 & $\times 10^{-25}$ \\
$S'(0)_{pp}$                                &  4. & 52 & $\times 10^{-24}$
             &  4. & 67 & $\times 10^{-24}$ &  4. & 52 & $\times 10^{-24}$ \\
$S(0)_{33}$                                 &  5. & 00 &
             &  5. & 00 &                   &  5. & 00 &                  \\
$S'(0)_{33}$                                & -0. & 9  &
             & -0. & 9  &                   & -0. & 9 &                  \\
$S(0)_{34}$                                 &  5. & 33 & $\times 10^{-4}$
             &  5. & 4  & $\times 10^{-4}$  &  5. & 33 & $\times 10^{-4}$ \\
$S'(0)_{34}$                                & -3. & 1  & $\times 10^{-4}$
             & -3. & 10 & $\times 10^{-4}$  & -3. & 10 & $\times 10^{-4}$ \\
$S(0)_{17}$                                 &  2. & 24 & $\times 10^{-5}$
             &  2. & 24 & $\times 10^{-5}$  &  2. & 24 & $\times 10^{-5}$ \\
$S'(0)_{17}$                                & -3. & 00 & $\times 10^{-5}$
             & -3. & 00 & $\times 10^{-5}$  & -3. & 00 & $\times 10^{-5}$ \\
\tableline
$S(0)_{{}^{12}\protect\text{C}+p}$          &  1. & 45 & $\times 10^{-3}$
             &  1. & 40 & $\times 10^{-3}$  &  1. & 40 & $\times 10^{-3}$ \\
$S'(0)_{{}^{12}\protect\text{C}+p}$         &  2. & 45 & $\times 10^{-3}$
             &  4. & 24 & $\times 10^{-3}$  &  4. & 24 & $\times 10^{-3}$ \\
$S(0)_{{}^{13}\protect\text{C}+p}$          &  5. & 50 & $\times 10^{-3}$
             &  5. & 50 & $\times 10^{-3}$  &  5. & 77 & $\times 10^{-3}$ \\
$S'(0)_{{}^{13}\protect\text{C}+p}$         &  1. & 34 & $\times 10^{-2}$
             &  1. & 34 & $\times 10^{-2}$  &  1. & 40 & $\times 10^{-2}$ \\
$S(0)_{{}^{14}\protect\text{N}+p}$          &  3. & 32 & $\times 10^{-3}$
             &  3. & 20 & $\times 10^{-3}$  &  3. & 32 & $\times 10^{-3}$ \\
$S'(0)_{{}^{14}\protect\text{N}+p}$         & -5. & 91 & $\times 10^{-3}$
             & -5. & 71 & $\times 10^{-3}$  & -5. & 91 & $\times 10^{-3}$ \\
$S(0)_{{}^{15}\protect\text{N}(p,\gamma){}^{16}\protect\text{O}}$
                                            &  6. & 40 & $\times 10^{-2}$
             &  6. & 40 & $\times 10^{-2}$  &  6. & 40 & $\times 10^{-2}$ \\
$S'(0)_{{}^{15}\protect\text{N}(p,\gamma){}^{16}\protect\text{O}}$
                                            &  3. & 00 & $\times 10^{-2}$
             &  3. & 00 & $\times 10^{-2}$  &  3. & 00 & $\times 10^{-2}$ \\
$S(0)_{{}^{15}\protect\text{N}(p,\alpha){}^{12}\protect\text{C}}$
                                            &  7. & 80 & $\times 10$
             &  5. & 34 & $\times 10$       &  7. & 04 & $\times 10$ \\
$S'(0)_{{}^{15}\protect\text{N}(p,\alpha){}^{12}\protect\text{C}}$
                                            &  3. & 51 & $\times 10^{2}$
             &\multicolumn{3}{c}{--}        &  4. & 21 & $\times 10^{2}$ \\
$S(0)_{{}^{16}\protect\text{O}+p}$          &  9. & 40 & $\times 10^{-3}$
             &  9. & 40 & $\times 10^{-3}$  &  9. & 40 & $\times 10^{-3}$ \\
$S'(0)_{{}^{16}\protect\text{O}+p}$         & -2. & 30 & $\times 10^{-2}$
             & -2. & 30 & $\times 10^{-2}$  & -2. & 30 & $\times 10^{-2}$ \\
\end{tabular}
\end{table}
\mediumtext
\begin{table}
\caption[dddd]{
   Comparison among recent Standard Solar Model
predictions and experimental results. For the definition of BP, TCL
and CDF94 see Table~\protect\ref{Inpar}.
We show the central temperature $T_c[10^7{}^o\protect\text{K}]$,
the Helium abundance in mass Y, the metallicity fraction Z, the values
of each component of the neutrino
flux $[10^9\protect\text{cm}^{-2}\protect\text{s}^{-1}]$, the calculated
signals for the Chlorine (Cl) and the Gallium (Ga) experiments [SNU].
On the right side we present the experimental constraints at one and
three standard deviation level.
   \label{SMflux}
               }
\begin{tabular}{lr@{}lr@{}lr@{}lr@{\ }r@{}lr@{\ }r@{}lc}
&\multicolumn{6}{c}{Standard Solar Models}
  &\multicolumn{7}{c}{Experimental Constraints} \\
&\multicolumn{2}{c}{BP}
  &\multicolumn{2}{c}{TCL}
     &\multicolumn{2}{c}{CDF94}
       &\multicolumn{3}{c}{$1\sigma$}
         &\multicolumn{3}{c}{$3\sigma$}
           & from \\
\tableline
$T_c$                               &   1. & 569
   &        1.  & 543               &   1. & 564
   &            &    &              &      &    &
   & \\
Y                                   &   0. & 273
   &        0.  & 271               &   0. & 289
   &            &    &              &      &    &
   & \\
Z ($\times 10^{2}$)                  &   1. & 96
   &        1.  & 88                &   1. & 84
   &            &    &              &      &    &
   & \\
\tableline
$pp$                                &  60. & 0
   &       60.  & 4                 &  60. & 0
   &            &    &              &      &    &
   & \\
$pep$                               &   0. & 14
   &        0.  & 14                &   0. & 14
   &            &    &              &      &    &
   & \\
$pp+pep$                            &  60. & 14
   &       60.  & 54                &  60. & 14
   & $\geq$     & 64.& 0            & $\geq$ & 61. & 0
   & Ga + Cl \\
$^7$Be                              &   4. & 89
   &        4.  & 25                &   4. & 79
   & $\leq$ & 0.  & 70              & $\leq$& 4. & 23
   & Ga + Cl \\
$^8$B ($\times 10^{3}$)             &   5. & 69
   &        4.  & 14                &   5. & 6
   & $\leq$&  3.  & 30              & $\leq$ & 4. & 10
   & Ka \\
$^{13}$N                            &   0. & 49
   &        0.  & 36                &   0. & 47
   &       &    &                  &           &     &
   &  \\
$^{15}$O                            &   0. & 43
   &        0.  & 30                &   0. & 40
   &            &    &                  &           &     &
   &  \\
$^{13}$N + $^{15}$O                 &   0. & 92
   &        0.  & 66                &   0. & 87
   & $\leq$& 0.  & 6                & $\leq$& 3. & 6
   & Ga + Cl \\
$^{17}$F ($\times 10^{3}$)          &   5. & 4
   &\multicolumn{2}{c}{--}          &   4. & 8
   &            &    &                  &           &     &
   &  \\
$hep$ ($\times 10^{6}$)             &   1. & 2
   &\multicolumn{2}{c}{--}          &   1. & 3
   &            &    &                  &           &     &
   &  \\
\tableline
Cl                                  &   8. & 0
   &        6.  & 1                 &   7. & 8
   & $\leq$ & 2. & 6                & $\leq$ & 3. & 0
   & Cl \\
Ga                                  & 132 &
   &      121  &                    & 130 &
  & $\leq$ & 88   &                 & $\leq$ & 108 &
   & Ga \\
\end{tabular}
\end{table}
\begin{table}
\caption[eeee]{
   Test of the homology relationships. We show
 $k=\langle T(m) / T(m)^{\protect\text{SSM}}\rangle$,
and its r.m.s. variation $\Delta k$ for several non-standard models,
obtained by varying: $s_{pp} = S_{pp}/ S_{pp}^{\protect\text{SSM}}$,
opa = opacity / opacity$^{\protect\text{SSM}}$,
$z = (Z/X) / (Z/X)^{\protect\text{SSM}}$, and
$t =$ age / age$^{\protect\text{SSM}}$.
The averages are performed over the mass shells in the region
{\bf (a)}  $M/M_0 \leq 0.3$, and {\bf (b)} $M/M_0 \leq 0.97$.
   \label{thomo}
               }
\begin{tabular}{dddddddddddd}
\multicolumn{12}{c}{{\bf (a)}} \\
\multicolumn{3}{c}{$S_{pp}$ variation} &
  \multicolumn{3}{c}{opacity variation} &
    \multicolumn{3}{c}{Z/X variation} &
      \multicolumn{3}{c}{age variation} \\
\tableline
\multicolumn{1}{c}{$s_{pp}$} &
 \multicolumn{1}{c}{$k$} &
  \multicolumn{1}{c}{$\Delta k$} &
     \multicolumn{1}{c}{opa} &
      \multicolumn{1}{c}{$k$} &
       \multicolumn{1}{c}{$\Delta k$} &
          \multicolumn{1}{c}{$z$} &
           \multicolumn{1}{c}{$k$} &
            \multicolumn{1}{c}{$\Delta k$} &
              \multicolumn{1}{c}{$t$} &
               \multicolumn{1}{c}{$k$} &
                \multicolumn{1}{c}{$\Delta k$} \\
\multicolumn{2}{c}{} &
 \multicolumn{1}{c}{[$10^{-3}$]} &
    \multicolumn{2}{c}{} &
     \multicolumn{1}{c}{[$10^{-3}$]} &
        \multicolumn{2}{c}{} &
         \multicolumn{1}{c}{[$10^{-3}$]} &
            \multicolumn{2}{c}{} &
             \multicolumn{1}{c}{[$10^{-3}$]} \\
\tableline
 1.25 & 0.976 & 1.36 & 0.9 & 0.988 & 0.72 & 0.5 & 0.962 &  1.46
                                          & 0.9 & 0.996 &  2.16 \\
 1.5  & 0.955 & 2.58 & 0.7 & 0.958 & 2.71 & 0.3 & 0.939 &  2.17
                                          & 0.7 & 0.987 &  6.56 \\
 1.75 & 0.939 & 3.04 & 0.6 & 0.939 & 3.71 & 0.2 & 0.925 &  2.40
                                          & 0.4 & 0.979 & 10.6 \\
 2.0  & 0.925 & 3.58 &     &       &      & 0.1 & 0.904 &  4.20
                                          & 0.2 & 0.971 & 14.8 \\
 2.5  & 0.902 & 4.45 &     &       &      &     &       &
                                          & 0.1 & 0.966 & 17.3 \\
 3.5  & 0.868 & 5.43 &     &       &      &     &       &
                                          &     &       &      \\
\tableline
\tableline
\multicolumn{12}{c}{{\bf (b)}} \\
\multicolumn{3}{c}{$S_{pp}$ variation} &
  \multicolumn{3}{c}{opacity variation} &
    \multicolumn{3}{c}{Z/X variation} &
      \multicolumn{3}{c}{age variation} \\
\tableline
\multicolumn{1}{c}{$s_{pp}$} &
 \multicolumn{1}{c}{$k$} &
  \multicolumn{1}{c}{$\Delta k$} &
     \multicolumn{1}{c}{opa} &
      \multicolumn{1}{c}{$k$} &
       \multicolumn{1}{c}{$\Delta k$} &
          \multicolumn{1}{c}{$z$} &
           \multicolumn{1}{c}{$k$} &
            \multicolumn{1}{c}{$\Delta k$} &
              \multicolumn{1}{c}{$t$} &
               \multicolumn{1}{c}{$k$} &
                \multicolumn{1}{c}{$\Delta k$} \\
\multicolumn{2}{c}{} &
 \multicolumn{1}{c}{[$10^{-3}$]} &
    \multicolumn{2}{c}{} &
     \multicolumn{1}{c}{[$10^{-3}$]} &
        \multicolumn{2}{c}{} &
         \multicolumn{1}{c}{[$10^{-3}$]} &
            \multicolumn{2}{c}{} &
             \multicolumn{1}{c}{[$10^{-3}$]} \\
\tableline
 1.25 & 0.977 &  1.82 & 0.9 & 0.989 & 1.07 & 0.5 & 0.962 &  2.80
                                           & 0.9 & 1.00  &  4.17 \\
 1.5  & 0.957 &  1.88 & 0.7 & 0.962 & 4.00 & 0.3 & 0.938 &  4.48
                                           & 0.7 & 1.00  & 12.8  \\
 1.75 & 0.941 &  5.00 & 0.6 & 0.944 & 5.17 & 0.2 & 0.923 &  5.98
                                           & 0.4 & 0.999 & 21.0 \\
 2.0  & 0.928 &  6.48 &     &       &      & 0.1 & 0.902 &  8.45
                                           & 0.2 & 1.00  &  2.97 \\
 2.5  & 0.905 & 10.3  &     &       &      &     &       &
                                           & 0.1 & 1.00  & 14.8 \\
 3.5  & 0.870 & 17.4  &     &       &      &     &       &
                                           &     &       &      \\
\end{tabular}
\end{table}
\narrowtext
\begin{table}
\caption[ffff]{
   The $\beta_i$ coefficients of the power laws that describe
the dependence of the neutrino fluxes on the temperature
($\Phi_i = \Phi_i^{\protect\text{SSM}}
( T_c / T_c^{\protect\text{SSM}} )^{\beta_i}$).
The components of neutrino flux that we consider
are shown in the first column.
The values presented are the best fit to the
numerical calculations performed when each input parameter is
varied in the range specified in the first row (same notation as
Table~\protect\ref{SMflux}).
   \label{betas}
               }
\begin{tabular}{lr@{}lr@{}lr@{}lr@{}l}
       &\multicolumn{2}{c}{$ s_{pp} $}
                  &\multicolumn{2}{c}{opa}
                             &\multicolumn{2}{c}{$ z $}
                                      &\multicolumn{2}{c}{$ t $} \\
       &\multicolumn{2}{c}{$ 1 \div 3.5 $}
                  &\multicolumn{2}{c}{$ 0.6 \div 1 $}
                             &\multicolumn{2}{c}{$ 0.1 \div 1 $}
                                      &\multicolumn{2}{c}{$ 0.1 \div 1 $} \\
\tableline
$pp$   & -0. & 60 & -0. & 63 & -0. & 73 & -0. & 85 \\
$^7$Be &  8. & 74 &  9. & 51 & 10. & 8  & 11. & 4  \\
$^{15}$N&15. & 1  & 12. & 0  & 30. & 9  &  8. & 58 \\
$^{16}$O&23. & 51 & 15. & 7  & 35. & 6  & 17. & 6  \\
$pep$  &  2. & 20 & -2. & 23 & -1. & 71 &  0. & 49 \\
$^8$B  & 22. & 3  & 20. & 76 & 21. & 5  & 20. & 2  \\
\end{tabular}
\end{table}
\begin{figure}
\caption[chain]{
   The $pp$ chain.
               }
\label{fig1}
\end{figure}
\begin{figure}
\caption[gafpp]{
   The Gallium signal $S_{\protect\text{Ga}}$
is shown as a function of the neutrino
flux $\Phi_{pp+pep}$. Standard neutrinos correspond to the area inside the
full (dot-dashed) lines if the Kamiokande (Chlorine) value for the
Boron contribution is used. The Gallium result $\pm 1 \sigma$ is shown. The
lower limit for the $pp$-I flux is thus
$\Phi_{pp+pep}^{\protect\text{min}} =
64.2  \times 10^9\protect\text{cm}^{-2}\protect\text{s}^{-1}$
($64.6  \times 10^9\protect\text{cm}^{-2}\protect\text{s}^{-1}$)
The upper limit
$\Phi_{pp+pep}^{\protect\text{max}} =
64.8  \times 10^9\protect\text{cm}^{-2}\protect\text{s}^{-1}$
is  given by the luminosity constraint.
                }
\label{fig2}
\end{figure}
\begin{figure}
\caption[gafbe]{
   The Gallium signal $S_{\protect\text{Ga}}$
is shown as a function of the neutrino flux
$\Phi_{\protect\text{Be}}$.
The Gallium result $\pm 1\sigma$ is shown (dashed lines). For
standard neutrinos, the allowed region is above the straight line
$S_{\text{Ga}}^{\text{min}}$.
The region consistent with the Gallium result and standard
neutrinos is the shaded area. The allowed flux has to be smaller than
$\Phi_{\text{Be}} =
 7  \times 10^8\protect\text{cm}^{-2}\protect\text{s}^{-1}$
($4  \times 10^8\protect\text{cm}^{-2}\protect\text{s}^{-1}$),
at $ 1 \sigma$, if the Boron contribution is
derived from Chlorine (Kamiokande) experiment. The result of our
SSM is also shown ($\Diamond$).
        }
\label{fig3}
\end{figure}
\begin{figure}
\caption[gafcno]{
   The Gallium signal $S_{\protect\text{Ga}}$
is shown as a function of the neutrino flux
$\Phi_{\protect\text{CNO}}$.
The Gallium result $\pm 1\sigma$ is shown (dashed lines). For
standard neutrinos, the allowed region is above the straight line
$S_{\text{Ga}}^{\text{min}}$.
The region consistent with the Gallium result and standard
neutrinos is the shaded area. The allowed flux has to be smaller than
$\Phi_{\text{CNO}} =
 6  \times 10^8\protect\text{cm}^{-2}\protect\text{s}^{-1}$
($2  \times 10^8\protect\text{cm}^{-2}\protect\text{s}^{-1}$),
at $ 1 \sigma$, if the Boron contribution is
derived from Chlorine (Kamiokande) experiment. The result of our
SSM is also shown ($\Diamond$).
        }
\label{fig4}
\end{figure}
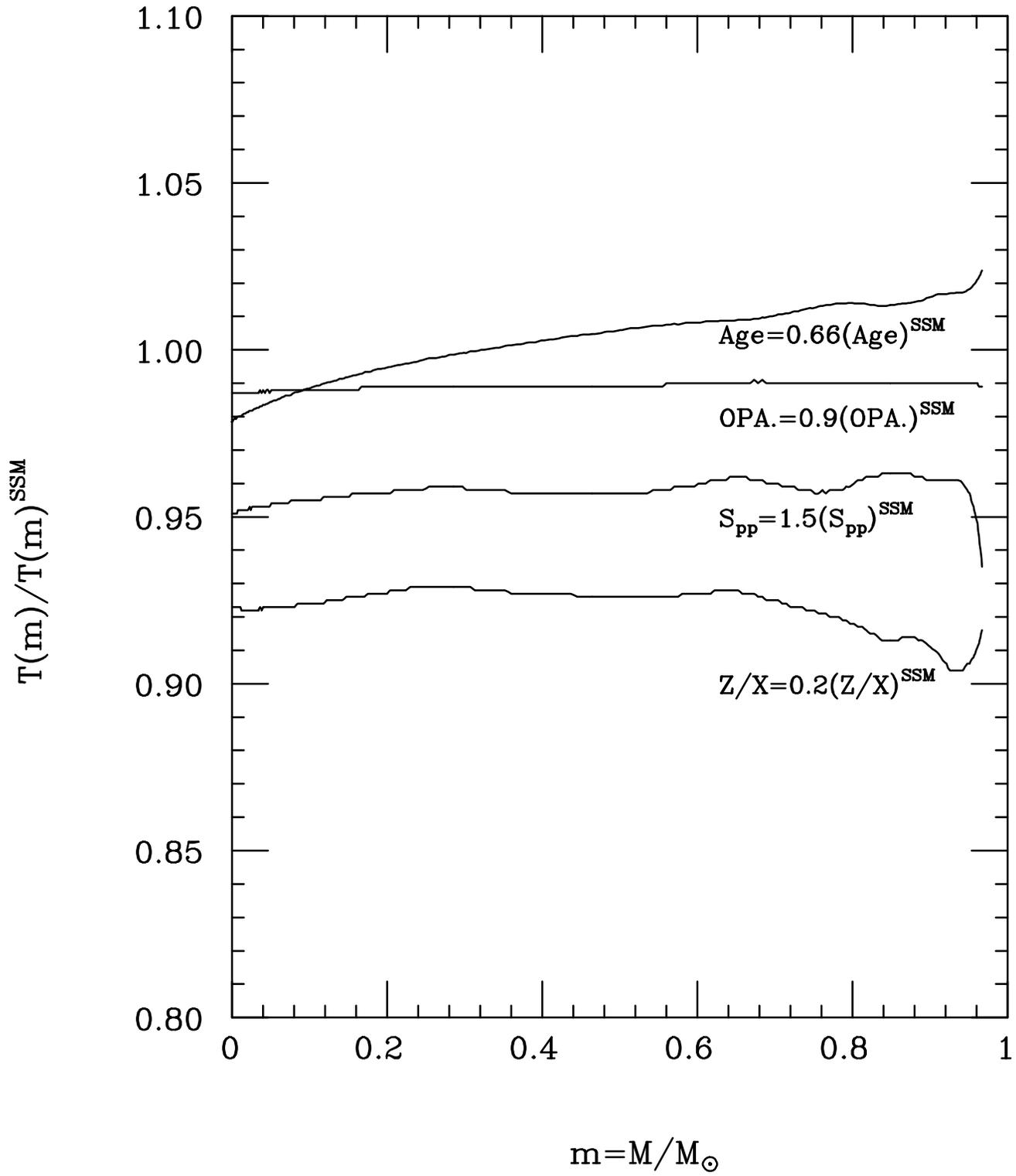
\begin{figure}
\caption[Temp]{
   The temperature profiles $T(m)$ normalized to
$T^{\protect\text{SSM}}(m)$ for a
few representative non-standard solar models.
        }
\label{fig5}
\end{figure}
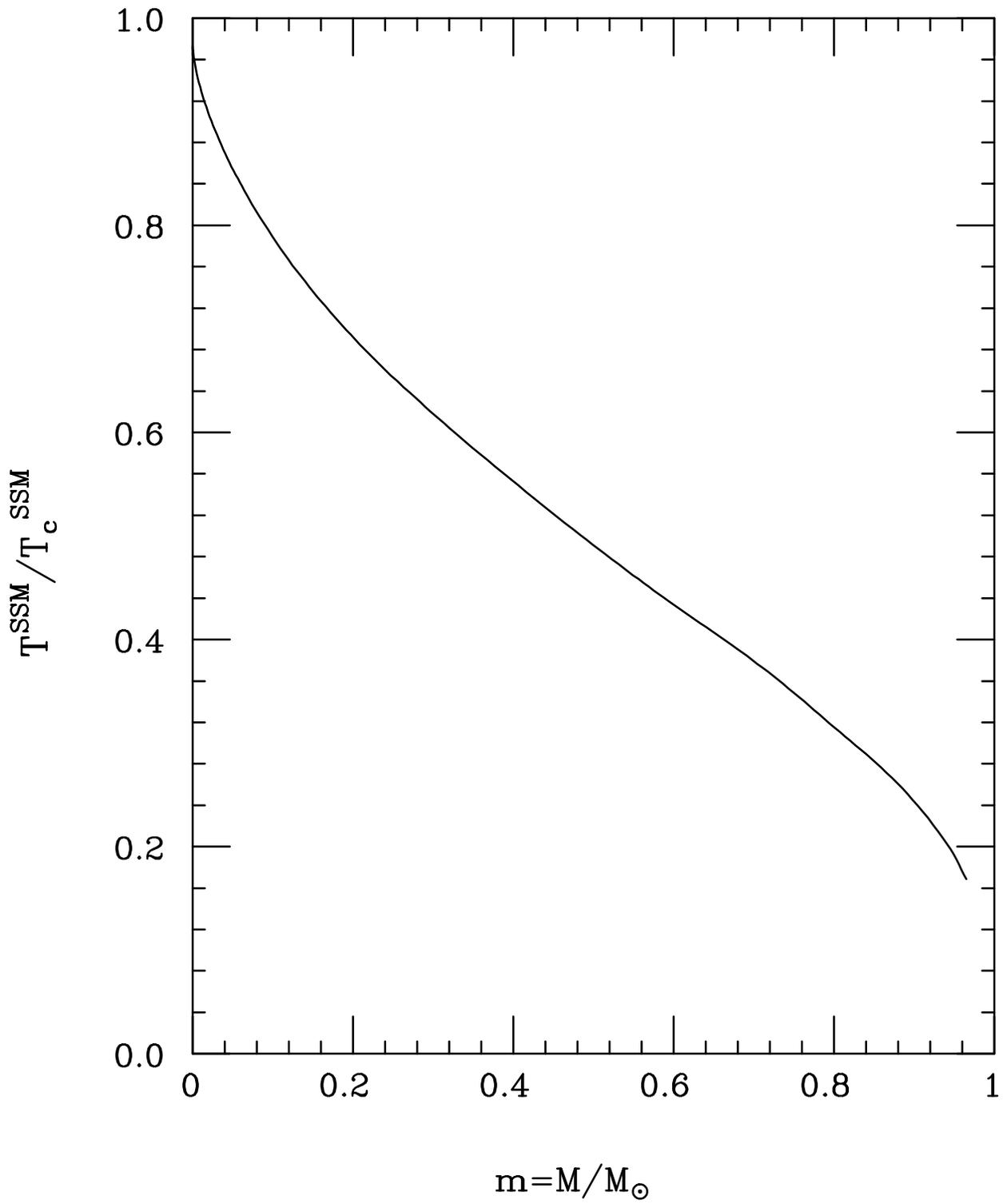
\begin{figure}
\caption[tp]{
The SSM temperature profile $T^{\protect\text{SSM}}(m)$,
normalized to the central value $T_c^{\protect\text{SSM}}$.
        }
\label{fig6}
\end{figure}

\begin{figure}
\caption[beha]{
The behaviour of
$\Phi_{pp}$, $\Phi_{\protect\text{Be}}$, and $\Phi_{\protect\text{B}}$
as a function of the central
temperature $T_c$ when varying $S_{pp}$,
opacity, Z/X and
age.
             }
\label{fig7}
\end{figure}

\begin{figure}
\caption[chit] {
 The $\chi^2$ as a function of the central temperature $T_c$.
(a) We use the standard value $S_{17}= 22.4$ eV barn~\protect\cite{John}.
 (b) We use the recently proposed value $S_{17}= 12$ eV
 barn~\protect\cite{Langa}.
        }
\label{fig8}
\end{figure}

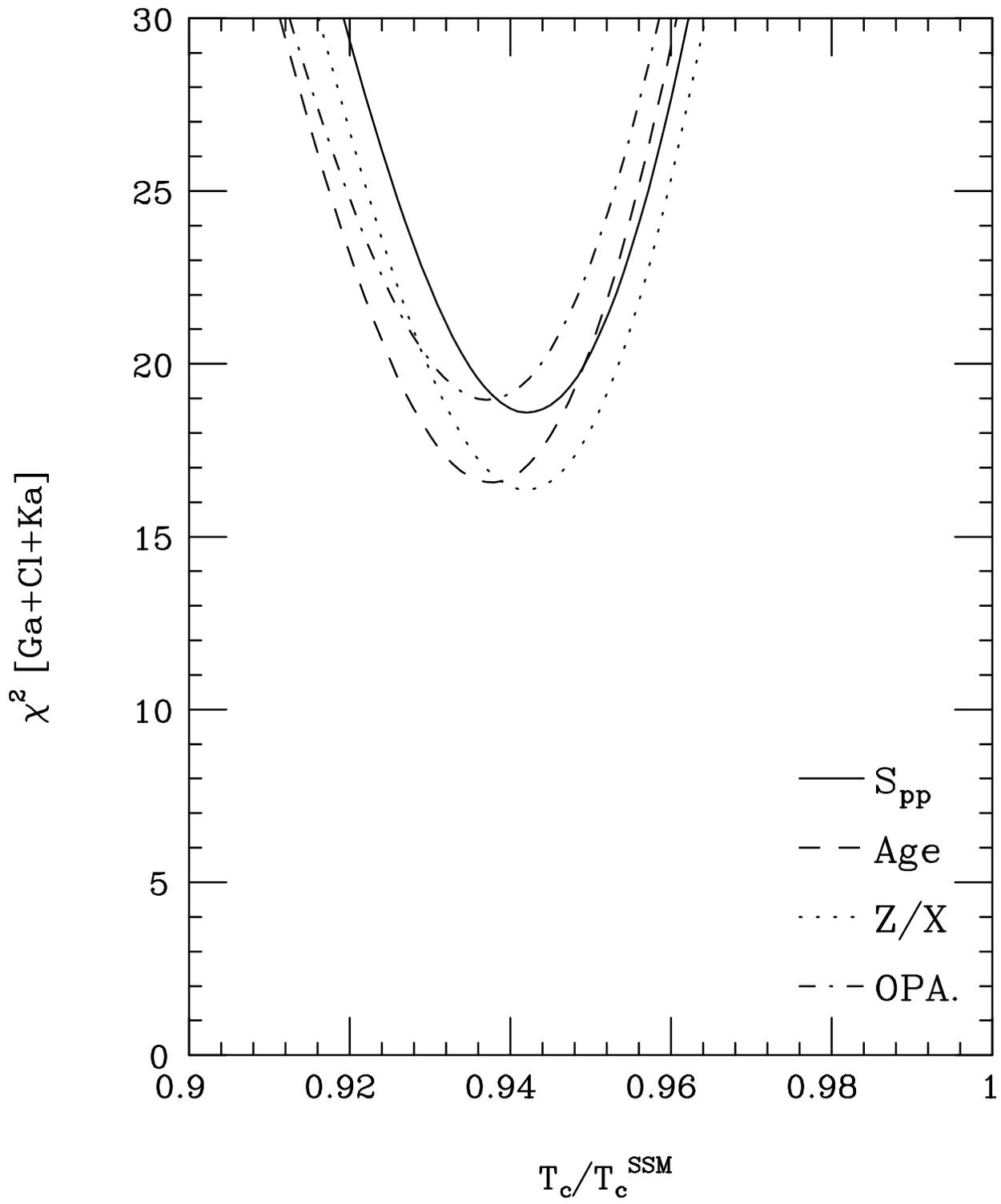
\begin{figure}
\caption[chipar]{
   The $\chi^2$ as a function of the central temperature $T_c$
when the
temperature variation is obtained by changing the different input
parameters.
        }
\label{fig9}
\end{figure}

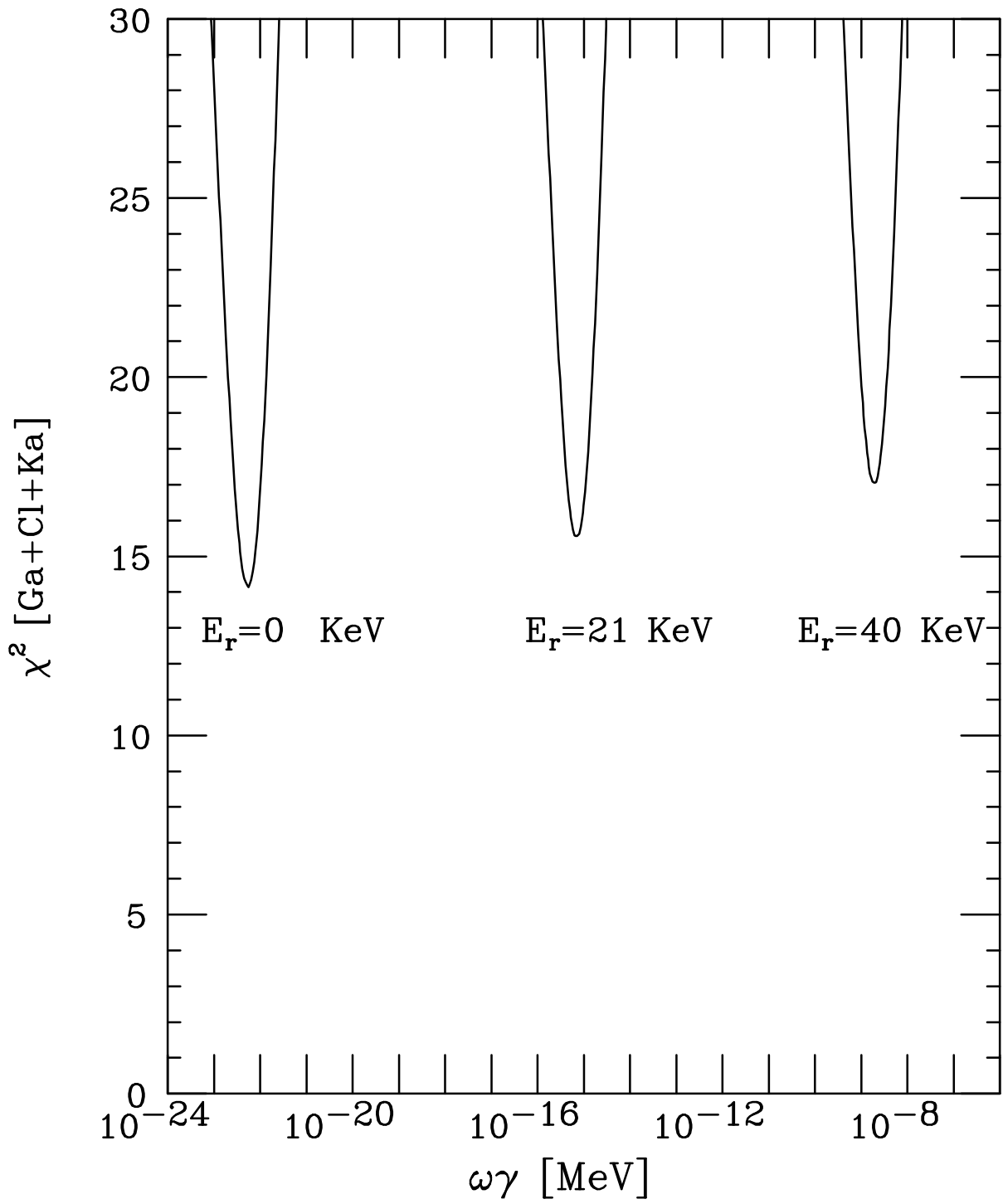
\begin{figure}
\caption[chir]{
   For a few values of the resonance energy $E_r$, we show $\chi^2$ as a
function of the resonance strength $\omega\gamma$.
The fit is done with all
(Ga+Cl+Ka) data.
        }
\label{fig10}
\end{figure}

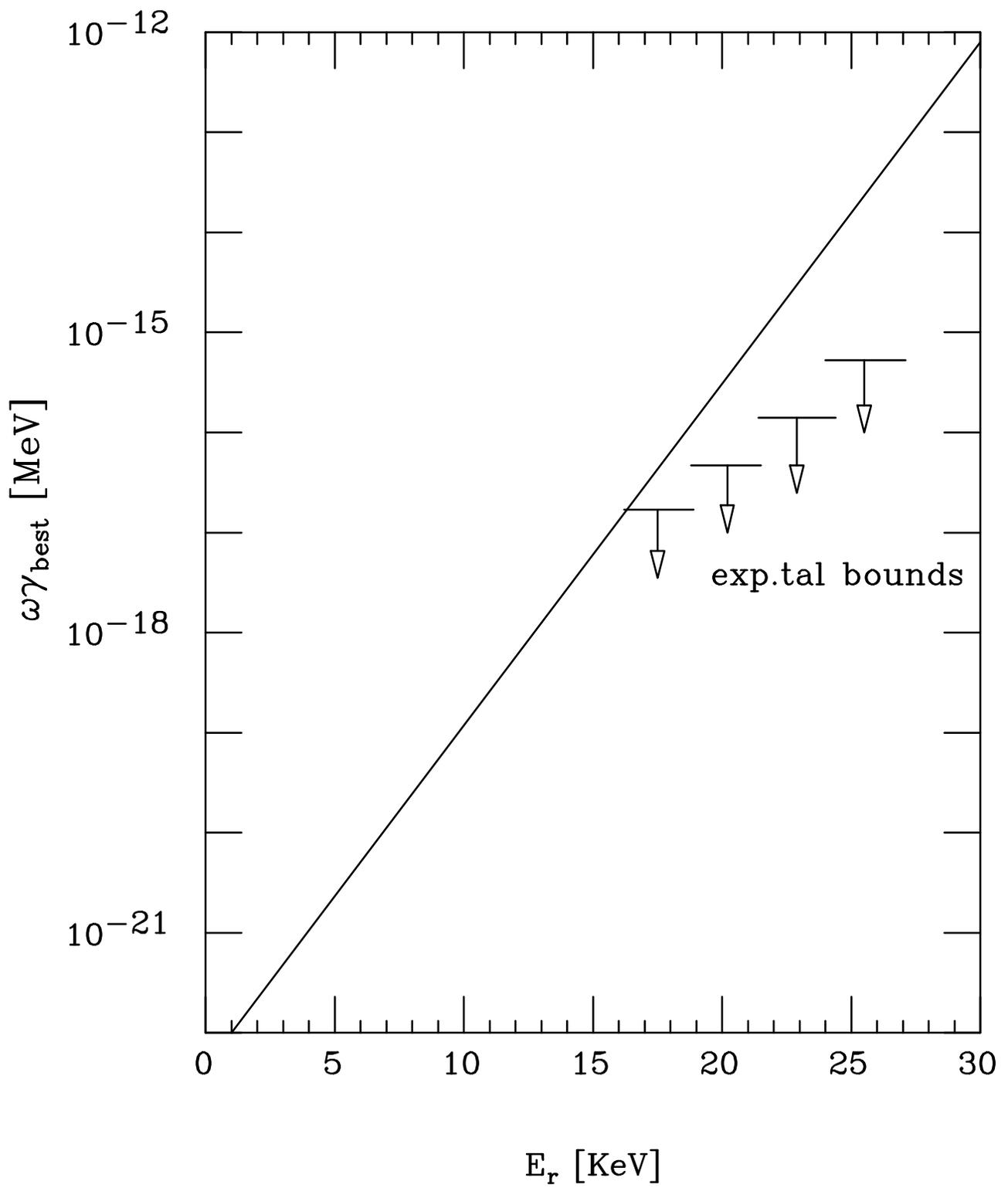
\begin{figure}
\caption[fitstrength]{
   The best fit strength $\omega\gamma$ of the
$^3\protect\text{He} + {}^3\protect\text{He}$ resonances as a
function of the resonance energy $E_r$ (full line). The arrows correspond
to the experimental upper bounds on the resonance strength, from
Ref.~\protect\cite{Krauss}.
        }
\label{fig11}
\end{figure}

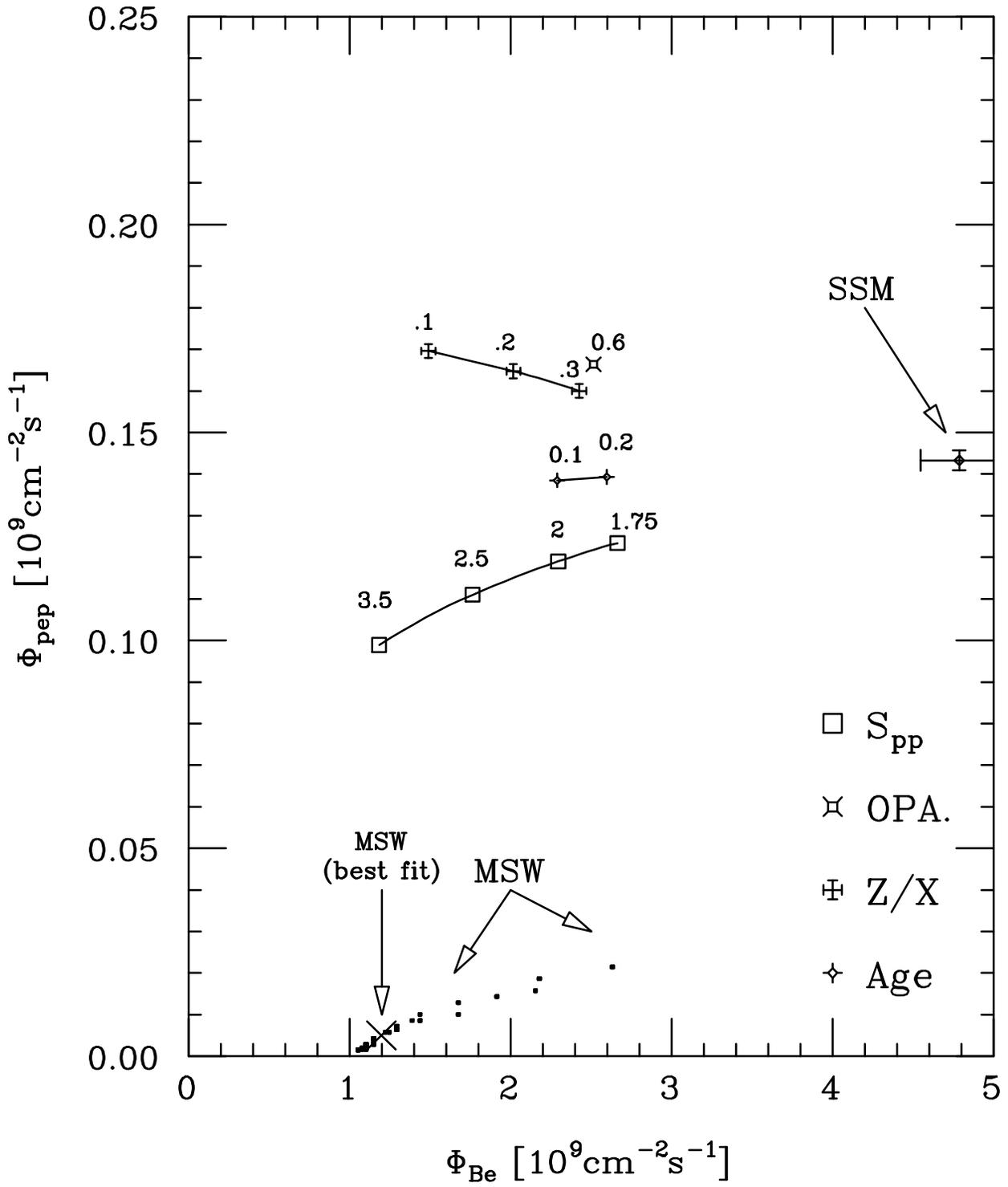
\begin{figure}
\caption[msw]{
   The $pep$ neutrino flux ($\Phi_{pep}$) vs. the
$^7$Be neutrino flux ($\Phi_{\protect\text{Be}}$).
For the standard solar model $(\Diamond)$.
For several non-standard solar models adjusted so as to reproduce
the Gallium result within $3 \sigma$ (the Boron contribution is
taken from the Kamiokande experiment); the notation is as in
Fig.~\protect\ref{fig7}, and
the number close to each point represents the corresponding value
of $\zeta=P/P^{\text{SSM}}$.
The values for the MSW solution, corresponding to the best fit $(\times)$,
and to the 90\% C.L. region (dots), see also Ref.~\protect\cite{Fiore}.
        }
\label{fig12}
\end{figure}
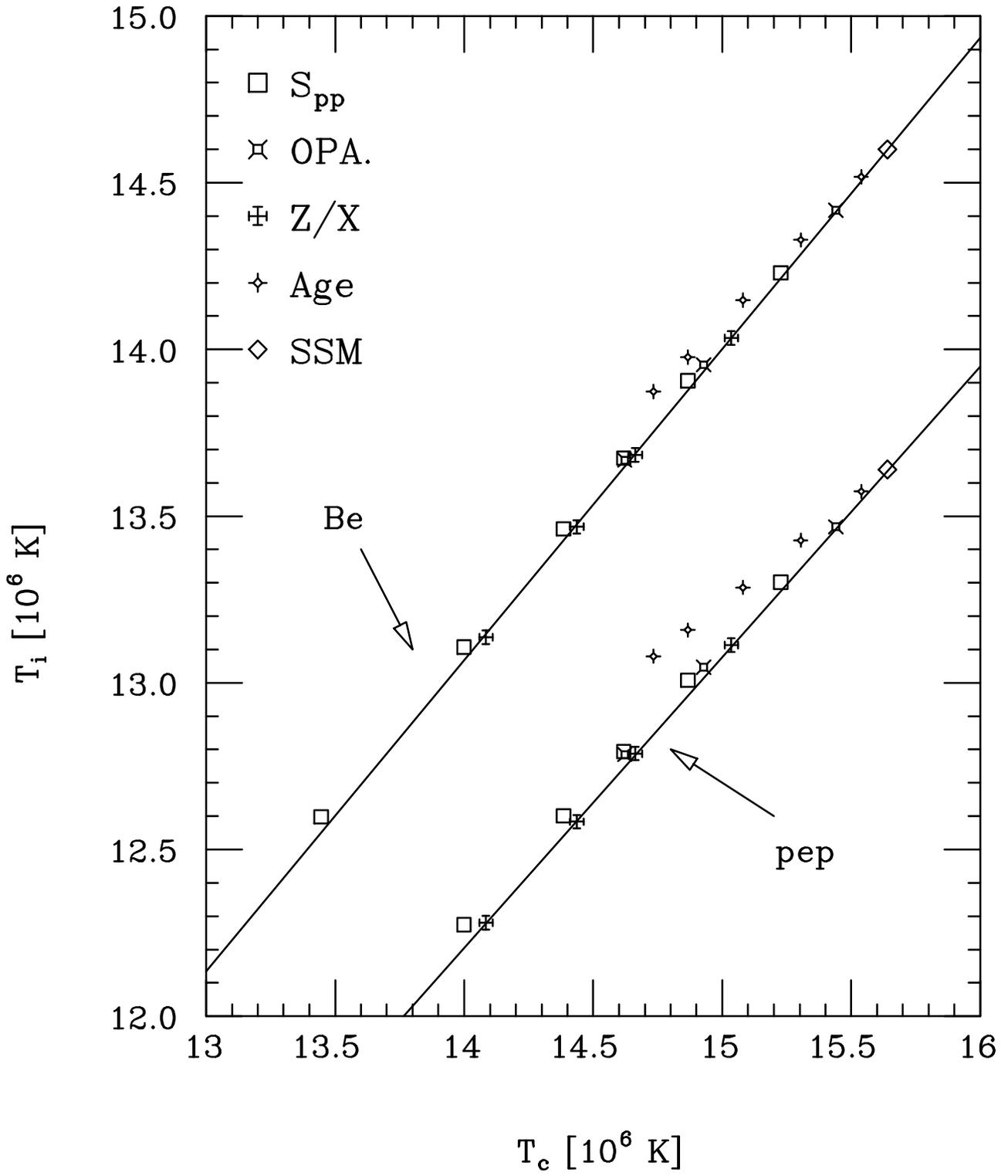
\begin{figure}
\caption[peak]{
  Relations among the temperatures $T_i$ at the $^7$Be and $pep$
peak production zones ($R/R_0=0.06$ and $R/R_0=0.09$, respectively), and
the central temperature $T_c$ in non-standard solar models. Data from
numerical calculations are shown with the same symbols as
in Fig.~\protect\ref{fig7}, while full lines show
the homology relations
$T_i= T_c \, ( T_i^{\text{SSM}}/T_c^{\text{SSM}})$.
        }
\label{fig13}
\end{figure}
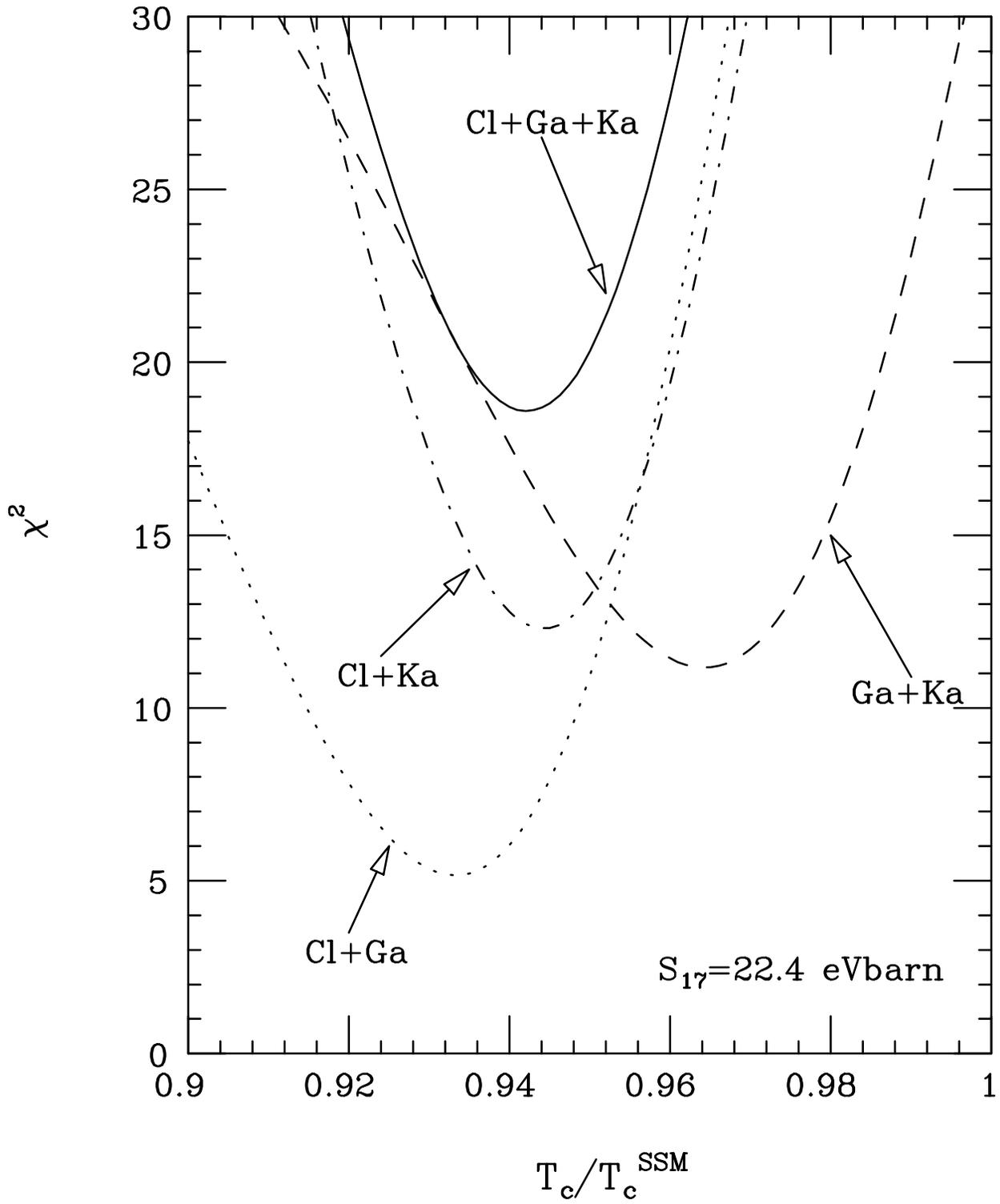
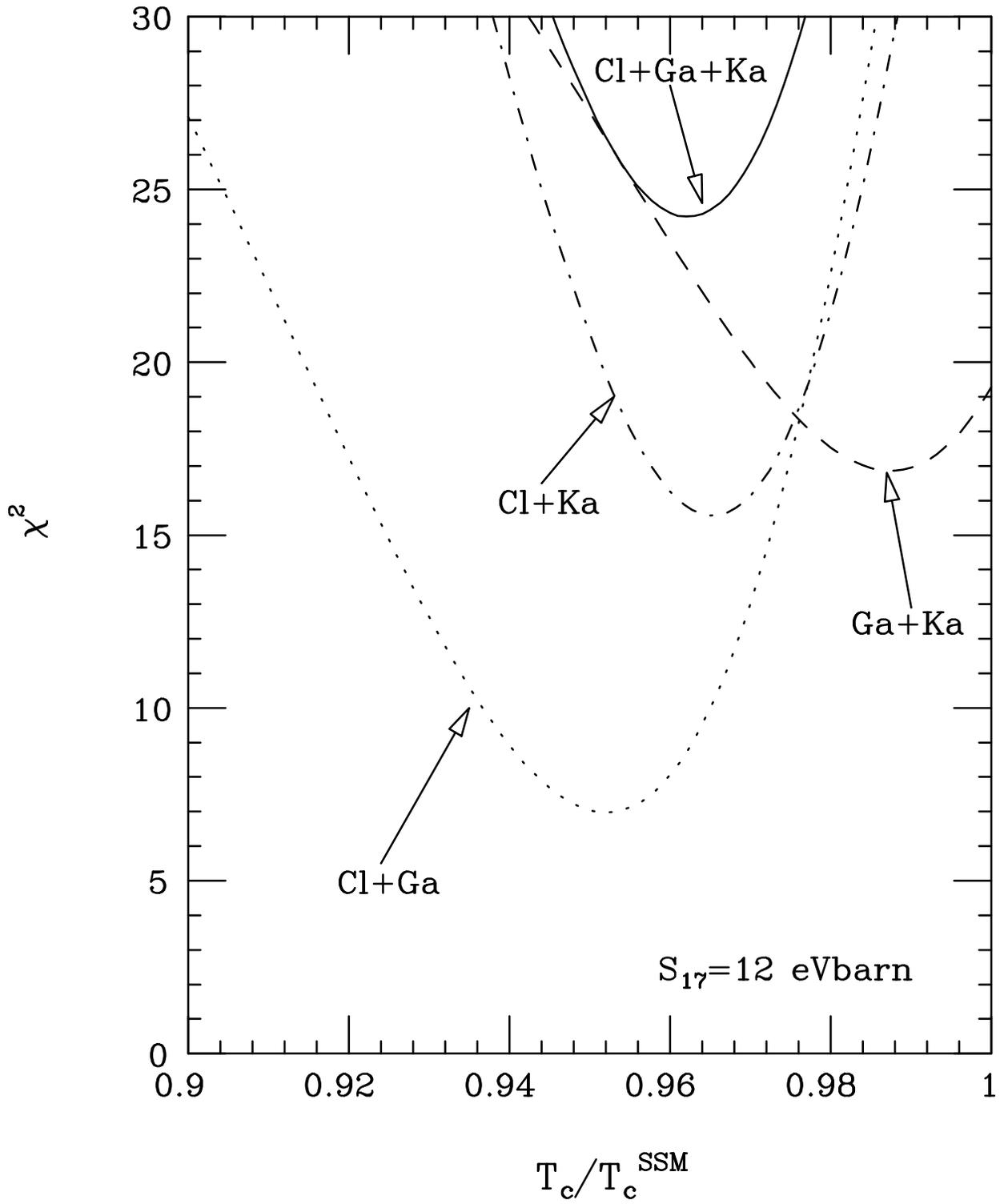
\end{document}